\documentclass[11.5pt]{article}
\usepackage{amsmath,amssymb}
\usepackage{url,lineno,microtype,subcaption}
\usepackage{hyperref}
\usepackage{amsthm}
\usepackage{multirow}
\usepackage[square,sort,comma,numbers]{natbib}
\usepackage{cancel}
\usepackage{graphicx}
\usepackage{color}

\newtheorem{thm}{Theorem}[section] 
\newtheorem{lemma}{Lemma}[section] 

\newtheorem{definition}{Definition}[section]
\newcommand{\bed}{\begin{definition}}
\newcommand{\eed}{\end{definition}}

\newcommand{\rom}[1]{\uppercase\expandafter{\romannumeral #1\relax}}

\newcommand{\eps}{\epsilon}
\usepackage{bbm}

\newcommand{\bitem}{\begin{itemize}}
\newcommand{\eitem}{\end{itemize}}

\newcommand{\goto}{\rightarrow}

\newcommand{\beqn}{\begin{equation}}
\newcommand{\eeqn}{\end{equation}}
\newcommand{\balign}{\begin{align}}
\newcommand{\ealign}{\end{align}}

\usepackage{amssymb}
\newcommand{\beq}{\begin{equation}}
\newcommand{\eeq}{\end{equation}}

\allowdisplaybreaks

\usepackage{enumitem}
\newcommand{\hamm}{\mathrm{Hamm}} 
 
\newcommand{\hell}{\hat{Y}}
\usepackage[normalem]{ulem}

\title{Subject clustering by IF-PCA and several recent methods}
\begin{document}
\author{Dieyi Chen$^*$,  Jiashun Jin$^\dagger$, and Zheng Tracy Ke$^*$ \\
Department of Statistics \\
Carnegie Mellon University$^\dagger$ and Harvard University$^*$} 
\date{}
\maketitle

\begin{abstract} 
 Subject clustering (i.e., the use of measured features to cluster subjects, such as patients or cells, into multiple 
groups)  is a problem of great interest.  In recent years, many approaches were proposed,  among which unsupervised deep learning (UDL) has 
received a great deal of attention. Two interesting questions are (a) how to combine 
the strengths of UDL and other approaches, and (b) 
how these approaches compare to one other.

We combine Variational Auto-Encoder (VAE), a popular UDL approach, with the recent idea of Influential Feature PCA (IF-PCA), and propose IF-VAE as a new method for subject clustering.  
We study IF-VAE and compare it with several other methods (including IF-PCA, VAE, Seurat, and SC3) 
on $10$ gene microarray data sets and $8$ single-cell RNA-seq data sets. 
We find that IF-VAE significantly improves over VAE, but 
still underperforms IF-PCA. 
We also find that IF-PCA is quite competitive, which slightly outperforms 
Seurat and SC3 over the $8$ single-cell data sets. 
IF-PCA is conceptually simple and permits delicate analysis. 
We demonstrate that IF-PCA is capable of achieving the phase transition in a Rare/Weak model.   
Comparatively, Seurat and SC3 are 
more complex and theoretically difficult to analyze (for these reasons, their optimality 
remains unclear). 
\end{abstract}

\section{Introduction}  
We are interested in the problem of {\it high-dimensional clustering} or {\it subject clustering}.  
Suppose we have a group of $n$ subjects (e.g.,  patients or cells)  measured on the 
same set of $p$ features (e.g., genes). The subjects 
come from $K$ different classes or groups (e.g., normal group and diseased group), 
but unfortunately, the class labels are unknown.  
In such a case, we say the data are {\it unlabeled}. 
For $1 \leq i \leq n$, denote the class label of subject $i$ by $Y_i$ and 
denote the $p$-dimensional measured feature vector of subject $i$ by $X_i$.  
Note that $Y_i$ take values from $\{1, 2, \ldots, K\}$.  The class labels are unknown and the goal is to predict them using the measured features $X_1, X_2, \ldots, X_n$. 

High-dimensional clustering is an unsupervised learning problem. It is especially interesting in the {\it Big Data era}: although the volume of available scientific data grows rapidly, a significant fraction of them are unlabeled. 
In some cases, it is simply hard to label each individual sample (e.g., action unit recognition \citep{AU}). 
In some other cases, labeling each individual sample is not hard, 
but due to the large sample size, it takes a huge amount of time and efforts to label the whole data set (e.g., ImageNet \citep{ImageNet}).  
In other instances (e.g., cancer diagnosis), we may have a preliminary opinion on how to label the data, but we are unsure of the labels' accuracy, so we would like a second, preferably independent, opinion.
In all these cases, we seek an effective and user-friendly clustering method.

In recent years, the area of high-dimensional clustering has witnessed exciting advancements in several directions.
First, many new types of data sets (e.g., 
sing-cell data) have emerged and become increasingly more accessible. 
Second, remarkable successes have been made on nonlinear modeling for high dimensional data,  
and several Unsupervised Deep Leaning (UDL) approaches have been proposed \citep{FanMa}, 
including but not limited to Variational Auto-Encoder (VAE) and Generative Adversarial Network (GAN).  
Last but not the least,   several clustering methods for single-cell data (e.g., Seurat \citep{Satija} and SC3 \citep{Vladimir})  
have been proposed and become popular.  

In this paper, we are primarily interested in Influential-Feature Principal Component Analysis (IF-PCA), 
a clustering algorithm proposed by \cite{JinWang}. 
As in many recent works in high-dimensional data analysis (e.g., \cite{ABDJ},   \cite{Paul}), 
we assume  
\begin{itemize} 
\item  $p \gg n \gg 1$ 
\item  out of all $p$ measured features, only a small fraction of them are relevant to clustering decision.  
\end{itemize} 
IF-PCA is easy-to-use and does not have  tuning parameters. It is conceptually simple, and (on a high-level) 
contains two steps as follows. 
\begin{itemize} 
\item {\it IF-step}. A feature selection step that selects a small fraction of measured features which 
we believe to be influential or significant to the clustering decision. 
\item {\it Clustering step}.  A clustering step in which PCA (as a spectral clustering approach) is applied to all retained features.
\end{itemize}  
Instead of viewing IF-PCA as a specific clustering algorithm, we can 
 view it as a {\it generic two-step clustering approach}: for each of the two steps, we can choose methods that may vary from occasion to occasion in order to best suit the nature of the data. We anticipate that IF-PCA will adapt and develop over time as new data sets and tasks emerge.

\cite{JinWang} compared IF-PCA to a number of clustering algorithms (including the classical kmeans \citep{kmeans}, kmeans++ \citep{kmeans+}, SpectralGem \citep{Lee}, hierarchical clustering \citep{HTF} and sparse PCA \citep{ZHT}) using $10$ microarray data sets. They found that IF-PCA was competitive in clustering accuracy. Later,   \cite{JinKeWang} developed a theoretical framework for clustering and showed that 
 IF-PCA is optimal in the Rare/Weak signal model (a frequently used model in high-dimensional data analysis (\cite{DJ04}, \cite{DJ15}).

These appealing properties of IF-PCA motivate a revisit of this method. Specifically, we are interested in the two questions listed below.  
\begin{itemize} 
\item   There are many recent clustering algorithms specifically designed for single-cell data, such as Seurat \citep{Satija}, SC3 \citep{Vladimir}, RaceID \citep{RaceID}, ACTIONet \citep{ACTIONet}, Monocle3 \citep{Monocle3}, and SINCERA \citep{SINCERA}. 
Also, many UDL algorithms have been proposed and become well-known in recent years.  
An interesting question is how IF-PCA compares with these popular algorithms. 

\vspace{0.3 em} 
\noindent 
\cite{JinWang} only examined IF-PCA on gene microarray data. 
The single-cell RNA-seq data 
are similar to gene microarray data in some aspects but also have some distinguished characteristics (e.g., singel-cell RNA-sequencing provides an unbiased view of all transcripts and is therefore reliable for accurately measuring gene expression level changes \citep{RNAseq}).  
How IF-PCA compares to other popular methods for subject clustering with single-cell data is an intriguing question.

\item The PCA employed in the clustering step of IF-PCA is a linear method. Although we believe that 
the associations between class labels and measured features may be nonlinear,  the significance of the nonlinear effects is unclear.   To investigate this, we may consider a variant of IF-PCA in which PCA is replaced by some non-linear UDL methods in the clustering step.  An interesting question is how this variant  compares to IF-PCA and standard UDL methods (which has no IF-step).   
It helps us understand how significant the nonlinear effects are.
\end{itemize} 
To answer these questions, first, we propose a new approach, IF-VAE, which combines the main idea of IF-PCA with the 
Variational Auto-Encoder (VAE) \citep{VAE}  (one of the most popular 
Unsupervised Deep Learning approaches in recent literatures). 

Second, we compare IF-VAE with several methods including VAE, IF-PCA, SpectralGem \citep{Lee}, and classical kmeans,   using the  $10$ microarray data sets in \cite{JinWang}. We find that 
\begin{itemize} 
\item Somewhat surprisingly, VAE underperforms most other methods, including the classical  kmeans.  
\item IF-VAE, whic combines VAE with the IF-step of IF-PCA, significantly outperforms VAE. 
\item The performance of IF-PCA and IF-VAE is comparable for approximately half of the data sets, whereas IF-VAE significantly underperforms IF-PCA for the remaining half of the data sets.
\end{itemize}  
These results suggest the following:  
\begin{itemize} 
\item (a).  The idea of combining the IF step in the IF-PCA with VAE is valuable. 
\item (b).  Deep neural network methods do not appear to have a clear advantage for this type of 
data sets. 
\end{itemize} 
For (b), one possible reason is that the associations between class labels and measured features 
are not highly nonlinear. Another possible reason is that existing deep neural network approaches need further improvements in order to perform satisfactorily on these data sets.   
Since IF-PCA and IF-VAE use the same IF-step, the unsatisfactory performance of IF-VAE is largely attributable to the VAE-step and not the IF-step.
To see this, we note that SpectralGem is essentially the classical PCA clustering method (see Section \ref{subsec:PCA}). 
VAE does not appear to show an advantage over SpecGem, explaining why IF-VAE cannot outperform IF-PCA. 

Last, we compare IF-VAE with IF-PCA, Seurat and SC3 on $8$ single-cell  RNA-seq data sets. 
We observe that
\begin{itemize} 
\item  IF-VAE continues to underperform other methods on the $8$ single-cell data sets, but similar as above, 
the unsatisfactory performance is largely attributable to the VAE step and not the IF-step. 
\item IF-PCA outperforms SC3 slightly and outperforms Seurat more significantly. 
\end{itemize} 
At the same time, we note that 
\begin{itemize} 
\item Seurat has four tuning parameters and is the method that has the shortest execution time. 
\item The idea of SC3 is quite similar to that of IF-PCA, except that SC3 has a ``consensus voting" step that aggregates the strengths of many clustering results. With consensus voting, SC3 may empirically perform more satisfactorily, but it is also more complex internally. Regarding the computational cost, it runs much slower than IF-PCA due to the consensus voting step.
\end{itemize} 
Moreover,  IF-PCA is conceptually simple and permits fine-grained analysis.   In Section \ref{sec:theory},  
we develop a theoretical framework and show that IF-PCA achieves the optimal phase transition 
in a Rare/Weak signal setting.  Especially, we show in the region of interest (where 
successful subject clustering is possible), 
\begin{itemize} 
\item if the signals are less sparse, signals may be individually weak.  In this case, PCA 
is optimal (and IF-PCA reduces to PCA if we choose the IF-step properly). 
\item if the signals are more sparse, the signals need to be relatively strong (so successful clustering is possible). 
In this case, feature selection is necessary, and IF-PCA is optimal. However, PCA 
may be non-optimal for it does not use a feature selection step.  
\end{itemize} 
In comparison, other popular methods are difficult to analyze theoretically, hence, their optimality is unclear. We note that hard-to-analyze methods will also be hard 
to improve in the future.

In conclusion, IF-PCA is quite competitive compared to the recently popular subject clustering methods, both for gene microarray data and single-cell data. It is worthwhile to study IF-PCA both theoretically and in (a variety of) applications. IF-VAE is a significant improvement over VAE, but it is still inferior to other prevalent methods in this area (the underperformance is largely due to the VAE step, not the IF-step). It is desirable to further improve IF-VAE (especially the VAE step) to make it more competitive.

\section{Models and methods} \label{sec:methods} 
As before, suppose  we have measurements on the same set of $p$ features for $n$ samples. 
Denote the data matrix by $X \in \mathbb{R}^{n, p}$, and write 
\begin{equation} \label{model1} 
X =  [X_1, X_2, \ldots, X_n]' = [x_1, x_2, \ldots, x_p],   
\end{equation} 
where $X_i \in \mathbb{R}^p$ denotes the measured feature vector for sample $i$,  $1 \leq i \leq n$.     
From time to time, we may want to normalize the data matrix before 
we implement any approaches.  
For $1 \leq j \leq p$, let $\hat{X}(j)$ and $\hat{\sigma}(j)$ be the empirical mean and standard deviation associated with feature $j$ (column $j$ of $X$), respectively.  
We normalize each column of $X$ and denote the resultant matrix by $W$, where 
\begin{equation} \label{normalization} 
W = [w_1, w_2, \ldots, w_p] = [W_1, W_2, \ldots W_n]' \in \mathbb{R}^{n, p},   \;\;\;  \mbox{and}  \;\;\;  W_i(j) = [X_i(j) - \hat{X}(j)]/ \hat{\sigma}(j).  
\end{equation} 
Below in Section \ref{subsec:models}, we introduce two models for $X$;  
then in Sections \ref{subsec:PCA}-\ref{subsec:SC3}, we describe 
the clustering methods considered in this paper, some of which (e.g., IF-VAE, IF-VAE(X),  IF-PCA(X)) 
are new.

\subsection{Two models}  
\label{subsec:models} 
A reasonable model is as follows.  We encode the class label $Y_i$ as a $K$-dimensional vector $\pi_i$, where 
$\pi_i = e_k$ if and only if sample $i$ belongs to class $k$,  and $e_k$ is the $k$-th standard 
Euclidean basis vector of $\mathbb{R}^K$, $1 \leq k \leq K$. Let $M = [\mu_1, \mu_2, \ldots, \mu_K]$ where 
$\mu_k \in \mathbb{R}^p$ is the mean vector for class $k$. We assume 
\begin{equation} \label{model1a}  
\mbox{$\mathbb{E}[X_i] = \mu_k$ if and only if subject $i$ belongs to class $k$}, \qquad  \mbox{or equivalently $
\mathbb{E}[X_i] = M \pi_i$}. 
\end{equation}  
Let $\Pi = [\pi_1, \pi_2, \ldots, \pi_n]'$ be the matrix of encoded class labels.  We can rewrite (\ref{model1a})  as 
\begin{equation} \label{model1b} 
X = \mathbb{E}[X] + (X - \mathbb{E}[X])  = ``\mbox{signal matrix}" + ``\mbox{noise matrix}",  
\qquad 
\mathbb{E}[X] = \Pi M'. 
\end{equation}
Also, it is reasonable to assume that out of many measured features, only a small fraction of them are
useful in the clustering decision. Therefore, letting $\bar{\mu} = (1/K) \sum_{k=1}^K \mu_k$, 
we assume 
\begin{equation} \label{model1c}  
\mbox{$\mu_1, \mu_2, \ldots, \mu_K$ are linearly independent  and $\mu_k - \bar{\mu}$ is sparse for each $1 \leq k \leq K$}. 
\end{equation} 
It follows that the $n \times p$ signal matrix $\mathbb{E}[X]$ has a rank $K$.  

Recall that $W$ is the normalized data matrix. Similar to (\ref{model1c}), we may decompose $W$ as the sum of a signal matrix and a noise matrix. But due to the normalization, the rank of the signal matrix is reduced to $(K-1)$.  

In Model (\ref{model1a})-(\ref{model1c}), 
$\mathbb{E}[X_i] = M \pi_i$, which is a  linear function of the encoded class label vectors $\pi_i$. 
For this reason, we may view Model (\ref{model1a})-(\ref{model1c}) as a linear model. 
In many modern applications, linear models may be inadequate, and we may prefer to use a nonlinear model 

The recent idea of neural network modeling provides a wide class of nonlinear models, 
which may be useful for our setting.  As an alternative to Model (\ref{model1a})-(\ref{model1c}), 
we may consider a neural network model as follows.  
In this model, we assume 
\begin{equation} \label{model2} 
Y_i = f(X_i,\theta),   \qquad i = 1, 2, \ldots, n, 
\end{equation} 
where $f(x, \theta)$ belongs to a class of nonlinear functions. For example, 
we may assume $f(x, \theta)$ belongs to the class of functions (without loss of generality, $x$ always includes a constant feature):
\[
\bigl\{f(x, \theta): f(x, \theta) = A_L (s_L (A_{L-1}  \ldots s_2 (A_2 s_1 (A_1 x))) | \theta = \{A_1, A_2, \ldots, A_L\}  \bigr\}, 
\] 
where $A_1, A_2, \ldots, A_L$ are matrices of certain sizes and $s_1, s_2, \ldots, s_L$ are some non-linear functions. 
Similar to Model (\ref{model1a})-(\ref{model1c}), we can impose some sparsity conditions on Model (\ref{model2}). See \cite{FanMa} for example.

\subsection{The PCA clustering approach and the SpectralGem} \label{subsec:PCA}
Principal Component Analysis (PCA) is a classical spectral clustering approach, which is especially appropriate  
for linear models like that in (\ref{model1a})-(\ref{model1c}) when the relevant features are non-sparse (see 
below for discussions on the case when the relevant features are sparse). 
The PCA clustering approach contains two simple steps as follows. 
Input: normalized data matrix $X$ and number of clusters $K$. Output: predicted class label vector $\hat{Y} = (\hat{Y}_1, \hat{Y}_2, \ldots, \hat{Y}_n)'$. 
\begin{itemize} 
\item Obtain the $n \times K$ matrix $\widehat{H} = [\hat{\eta}_1, \ldots, \hat{\eta}_K]$, 
where $\hat{\eta}_k$ is the $k$-th left singular vector of $X$ (associated with the $k$-th largest 
singular value of $X$). 
\item Cluster the $n$ rows of $\widehat{H}$ to $K$ groups by applying the classical kmeans assuming there are $\leq K$ classes. Let 
$\hat{Y}_i$ be the estimated class label of subject $i$. Output $\hat{Y}_1, \ldots, \hat{Y}_n$.  
\end{itemize} 

From time to time, we may choose to apply the PCA clustering approach to the normalized data matrix $W$.  
As explained before, we can similarly write $W$ as the sum of a  ``signal" matrix and a ``noise" matrix as in 
(\ref{model1c}), but due the normalization,  the rank of the ``signal" matrix under Model (\ref{model1a}) is reduced from $K$ to $(K-1)$. In such a case, we replace the $n \times K$ matrix $\widehat{H}$ by the $n \times (K-1)$ matrix 
\[
\widehat{\Xi} = [\hat{\xi}_1, \hat{\xi}_2, \ldots, \hat{\xi}_{K-1}], 
\] 
where similarly $\hat{\xi}_k$ is the $k$-th left singular vector of $W$.  

The PCA clustering approach has many modern variants, including but not limited to the SpectralGem \citep{Lee} and SCORE 
\citep{SCORE, SCORENormalization}. In this paper, we consider SpectralGem but skip the discussion 
on SCORE (SCORE was motivated by unsupervised learning in network and text data and shown to be effective on those types of data; it is unclear if SCORE is also effective for genetic and genomic data). 
Instead of applying PCA clustering to the data matrix $X$ (or $W$) directly, 
SpectralGem constructs an $n \times n$ symmetric matrix $M$, where 
$M(i,j)$ can be viewed as a similarity metric between subject $i$ and subject $j$.   
The remaining part of the algorithm has many small steps, but the 
essence is to apply the PCA clustering approach to the Laplacian normalized graph induced by 
$M$.   

The PCA spectral clustering approach is based on two important assumptions.    
\begin{itemize} 
\item The signal matrix $\mathbb{E}[X]$ is a linear function of class labels. 
\item It is hard to exploit sparsity in the data: either the data are non-sparse (such as the classical setting of $p \ll n$) or how to exploit sparsity is unclear.  
\end{itemize} 
In many modern settings, these assumptions are not satisfied:  the relationship between the signal matrix $\mathbb{E}[X]$ and class labels may be nonlinear, and it is highly desirable to exploit sparsity by adding a feature selection before conducting PCA clustering.    In such cases, 
we need an alternative approach. 
Below, we address respectively the non-linearity by VAE and 
the feature selection by IF-PCA.

\subsection{The Variational AutoEncoder (VAE) and VAE(X) clustering approaches}
\label{subsec:VAE} 
Given an $n \times p$ data matrix $X$ and an integer $d \leq  \mathrm{rank}(X)$, 
the essence of the PCA spectral clustering approach is to obtain a rank-$d$ approximation of $X$ is to use Singular Value Decomposition (SVD), 
\[
\widehat{X} =   \sum_{k=1}^d \sigma_k u_k v_k'.  
\] 
Here $\sigma_k$ is the $k$-th smallest singular value of $X$, and $u_k$ and $v_k$ are the corresponding left and right singular vectors of $X$, respectively.  Variational AutoEncoder (VAE) can be viewed as 
an extension of SVD, which obtains a rank-$d$ approximation 
of $X$ from training a neural network. The classical 
SVD is a linear method, but the neural network approach can be highly nonlinear.

VAE was first introduced by \cite{VAE} and has been 
 successfully applied to many application areas (e.g., image processing \citep{image-VAE},  computer vision \citep{CV-VAE},  and text mining \citep{text-VAE}).  
VAE consists of an encoder, a decoder, and a loss function.  
Given a data matrix $X \in \mathbb{R}^{n, p}$, the encoder embeds $X$ into a matrix $\widehat{Z} \in \mathbb{R}^{n, d}$   (usually $d \ll p$), and the decoder maps $\widehat{Z}$ back to the original data space and outputs a 
 matrix $\widehat{X} \in \mathbb{R}^{n, p}$,  which can be viewed as a rank-$d$ approximation 
of $X$. Different from classical SVD, $\widehat{X}$ is obtained in a nonlinear fashion by 
minimizing an objective that measures the information loss between $X$ and $\widehat{X}$.

A popular way to use VAE for subject clustering is as follows \citep{VAE-cluster}. 
Input: normalized data matrix $W = [w_1, w_2, \ldots, w_p] = [W_1, W_2, \ldots, W_n]'$, number of classes $K$,  dimension of the latent space $d$ (typically much smaller than $\min\{n, p\}$).  Output: predicted class label vector $\hat{Y} = (\hat{Y}_1, \hat{Y}_2, \ldots, \hat{Y}_n)$. 
\begin{itemize}
\item ({\it Dimension reduction by VAE}).  Train VAE and use the trained encoder to get an $n \times d$ matrix $\widehat{Z}$. 
\item ({\it Clustering}). Cluster all $n$ subjects into $K$ classes by applying k-means to the rows of 
$\widehat{Z}$.   Let $\hat{Y}$ be the predicted label vector.
\end{itemize}
Except for using a nonlinear approach to dimension reduction, 
VAE is similar to the PCA approach in clustering. 
We can apply VAE either to the normalized data matrix $W$ or the 
unnormalized data matrix $X$. We call them VAE(W) and  
VAE(X), respectively. In the context of using these notations, it is unnecessary to keep (W) and (X) at the same time, so we write 
VAE(W) as VAE for short (and to avoid confusion, we still write VAE(X) as VAE(X)).

\subsection{The orthodox IF-PCA and its variant IF-PCA(X)}
\label{subsec:IF-PCA} 
For many genomic and genetic data, Model (\ref{model1a})-(\ref{model1c}) is already a
reasonable model. We recall that under this model the normalized data matrix can be approximately written as 
\[
W = Q + (W-Q) =  ``\mbox{signal matrix}" + ``\mbox{noise matrix}", 
\] 
where approximately,  
\[
Q = \Pi [\mu_1 - \bar{\mu}, \mu_2 - \bar{\mu}, \ldots, \mu_K - \bar{\mu}]'  \in \mathbb{R}^{n, p},   
\] 
and is sparse (in the sense that only a small fraction of 
the columns  of $Q$ have a large $\ell^2$-norm; the $\ell^2$-norm of other columns are small or $0$). In such a setting, it is appropriate to conduct features selection, which removes 
a large amount of noise while keeping most nonzero columns of $Q$. 

Such observations motivate the (orthodox) IF-PCA.    The IF-PCA was first proposed in \cite{JinWang} and shown to have appealing clustering results on $10$ gene microarray data sets. 
In \cite{JinKeWang}, it was shown that IF-PCA 
is optimal in high-dimensional clustering. 
IF-PCA contains an IF step and a PCA step, and the IF-step contains 
two important components which we now introduce. 

The first component of the IF-step is the use of the 
Kolmogorov-Smirnov (KS) test for feature selection. Suppose we have $n$ (univariate) samples $z_1, z_2, 
\ldots, z_n$  from a cumulative distribution function (CDF) denoted by $F$. Introduce the empirical CDF by  
\begin{equation}  \label{KS1} 
F_n(t) = (1/n) \sum_{i = 1}^n 1\{z_i \leq t\}. 
\end{equation} 
Let $z = (z_1, z_2, \ldots, z_n)$.   The KS testing score is then 
\begin{equation} \label{KS2} 
\phi_n(z)  = \sqrt{n} \sup_{t} \{\|F_n(t) - F(t)\|\}. 
\end{equation} 
In the IF-PCA below, we take $F$ to be the theoretical CDF of $(z_i  - \bar{z}) / \hat{\sigma}$,   
where $z_i \stackrel{iid}{\sim} N(0,1)$, $1 \leq i \leq n$, and $\bar{z}$ and $\hat{\sigma}$ are the 
empirical mean and standard deviation of $z_1, z_2, \ldots, z_n$, respectively.

The second component  of the IF step is {\it Higher Criticism Threshold (HCT)}.  Higher Criticism was initially introduced by \cite{DJ04} (see also \cite{DJ15, ACV, Jager, HJ}) as a method for global testing. It has been recently applied to genetic data (e.g., \cite{barnett2017generalized}). 
HCT adapts Higher Criticism to a data-driven threshold choice \citep{JinWang}. It takes as input 
$p$ marginal $p$-values, one for a feature, and outputs a threshold for feature selection.  
Suppose 
we have $p$-values $\pi_1, \pi_2, \ldots, \pi_p$.   We sort them in the ascending order: 
\[
\pi_{(1)} < \pi_{(2)} < \ldots < \pi_{(p)}. 
\] 
Define the feature-wise HC score by $HC_{p, j} = \sqrt{p}(j/p - \pi_{(j)})/ \sqrt{\max\{\sqrt{n}(j/p - \pi_{(j)}),0 \} + j/p }$. The HCT is then 
\begin{equation} \label{HCT} 
\hat{t}_{HC} = \pi_{(\hat{j})}, \qquad \mbox{where $\hat{j} = \mathrm{argmax}_{ \{j: \pi_{(j)} > \log{p}/p,\, j < p/2 \}} \{HC_{p,j}\}$}. 
\end{equation}

IF-PCA runs  as follows. \\
Input: normalized feature vectors $W = [w_1, w_2, \ldots, w_p] = [W_1, W_2, \ldots, W_n]'$, 
number of classes $K$. Output: predicted class label vector $\hat{Y} = (\hat{Y}_1, \hat{Y}_2, \ldots, \hat{Y}_n)'$.  
\begin{itemize} 
\item (IF-step). For each $1 \leq j \leq p$,  compute a KS-score for feature $j$  by applying (\ref{KS1})-(\ref{KS2}) with $z = w_j$. Denote the KS scores by $\phi_{n}(w_1), \ldots, \phi_n(w_p)$ and let $\mu^*$ and $\sigma^*$ be their empirical mean and standard deviation, respectively.     Let $\psi_j^* = [\phi_{n}(w_j) - \mu^*] /\sigma^*$.  Compute the $p$-values by $\pi_j = 1 - F(\psi_{j}^*)$, where $F$ is the same CDF used in (\ref{KS2}). 
Obtain the HCT by applying (\ref{HCT}) to $\pi_1, \pi_2, \ldots, \pi_p$. Retain feature $j$ if $\pi_j \leq \hat{t}_{HC}$, and remove it otherwise.   
\item (Clustering-step).  Let $W^{IF}$ be the $n \times m$ sub-matrix of $W$ consisting 
of columns of $W$ corresponding to the retained features only  ($m$ is the number of retained features in (a)).  
For any $1 \leq k \leq \min\{m, n\}$, let  $\hat{\xi}_k^{IF}$ be the  left singular vector of $W^{IF}$  corresponding to the $k$-th largest singular value of $W^{IF}$. Let 
$\widehat{\Xi}^{IF} = [\hat{\xi}_1^{IF}, \ldots, \hat{\xi}_{K-1}^{IF}]  \in \mathbb{R}^{n, K-1}$.  
Cluster all $n$ subjects by applying the $k$-means to the $n$ rows of $\widehat{\Xi}^{IF}$, assuming there are $K$ clusters. Let $\hat{Y} = (\hat{Y}_1, \hat{Y}_2, \ldots, \hat{Y}_n)'$ be the predicted class labels. 
\end{itemize}  
In the IF-step, the normalization of $\psi_j^* = [\phi_{n}(w_j) - \mu^*] /\sigma^*$ is called Efron's null correction \citep{Efron}, a simple idea that is proved to be both necessary and effective for  analyzing genomic and genetic data \citep{JinWang2}.  We remark that although IF-PCA is motivated by the linear model in (\ref{model1c}), it is not tied to (\ref{model1c}) and 
is broadly applicable. In fact, the algorithm does not require any knowledge of Model (\ref{model1a})-(\ref{model1c}).  

In the (orthodox) IF-PCA, we apply both the IF-step and the clustering-step to the normalized data matrix $W$. 
Seemingly, for the IF-step, applying the algorithm to $W$ instead of the un-normalized data matrix $X$ is 
preferred. However, for the clustering-step, whether we should apply the algorithm to $W$ or $X$ remains unclear. 
We propose a small variant of IF-PCA by applying the IF-step and the clustering step to $W$ and $X$, respectively. 
\begin{itemize} 
\item (IF-step).  Apply exactly the same IF-step to $W$ as in the (orthodox) IF-PCA above.  
\item (Clustering-step).  Let $X^{IF}$ be the $n \times m$ sub-matrix of $X$ consisting 
of columns of $X$ corresponding to the retained features in the IF-step only.    
For any $1 \leq k \leq \min\{m, n\}$, let  $\hat{\eta}_k^{IF}$ be the  left singular vector of $X^{IF}$  corresponding to the $k$-th largest singular value of $X^{IF}$. Let 
$\widehat{H}^{IF} = [\hat{\eta}_1^{IF}, \ldots, \hat{\eta}_{K-1}^{IF}]  \in \mathbb{R}^{n, K-1}$.  
Cluster all $n$ subjects by applying the $k$-means to the $n$ rows of $\widehat{H}^{IF}$, assuming there are $K$ clusters. Let $\hat{Y} = (\hat{Y}_1, \hat{Y}_2, \ldots, \hat{Y}_n)'$ be the predicted class labels. 
\end{itemize} 
To differentiate from the (orthodox) IF-PCA (which we call IF-PCA below), 
we call the above variant IF-PCA(X).  See Table \ref{tab:method-sum} in Section \ref{subsec:method-sum}.  
The new variant was never proposed or studied before. 
It outperforms the (orthodox) 
IF-PCA in several data sets (e.g., see Section \ref{sec:result}). 

\subsection{IF-VAE and IF-VAE(X)}
\label{subsec:IF-VAE} 
Near the end of Section \ref{subsec:PCA}, we mention that the classical PCA has two disadvantages,  
not exploiting sparsity in feature vectors and not accounting for possible nonlinear relationships between 
the signal matrix and class labels. In Sections~\ref{subsec:VAE}-\ref{subsec:IF-PCA}, we have seen that VAE aims to exploit nonlinear relationships, and 
IF-PCA aims to exploit sparsity. We may combine VAE with the 
IF-step of IF-PCA for a simultaneous exploitation of sparsity and non-linearity. To this end, we propose a new algorithm called 
IF-VAE.  

IF-VAE contains an IF-step and a clustering step, and  runs as follows. 
Input: normalized data matrix $W = [w_1, w_2, \ldots, w_p] = [W_1, W_2, \ldots, W_n]'$, number of classes $K$,  dimension of the latent space in VAE (denoted by $d$).    Output: predicted class label vector $\hat{Y} = (\hat{Y}_1, \hat{Y}_2, \ldots, \hat{Y}_n)$. 
\begin{itemize} 
\item ({\it IF-step}). Run the same IF-step as in Section \ref{subsec:IF-PCA}, and let $W^{IF} = [W_1^{IF},\ldots, W_n^{IF}]' \in \mathbb{R}^{n \times m}$ be the matrix consisting of the retained features only (same as in the IF-step in IF-PCA, $m$ 
is the number of retained features).  
\item ({\it Clustering-step}).  Apply VAE with $W^{IF} \in \mathbb{R}^{n \times m}$ and 
obtain an $n \times d$ matrix $\widehat{Z}^{IF}$, which can be viewed as an estimation of the low-dimensional representation of $W^{IF}$.   Cluster the $n$ samples into $K$ clusters by applying the classical k-means to $\widehat{Z}^{IF}$ assuming there are $K$ classes. Let $\hat{Y}$ be the predicted label vector.
\end{itemize} 
In the clustering-step, we apply VAE to the normalized data matrix $W$. 
Similarly as in Section \ref{subsec:IF-PCA}, if 
we apply VAE to the un-normalized data matrix $X$, 
then we have a variant of IF-VAE, which we denote by IF-VAE(X). 
See Table \ref{tab:method-sum} in Section \ref{subsec:method-sum}.

\subsection{Seurat and SC3} 
\label{subsec:SC3} 
We now introduce Seurat and SC3,  two recent algorithms that are especially popular for 
subject clustering with Single-cell RNA-seq data. We discuss them separately. 

Seurat was proposed in \cite{Satija}.  On a high level, Seurat is quite similar to IF-PCA, 
and we can view it as having only two main steps: a feature selection step and 
a clustering step. But different from IF-PCA, Seurat uses a different feature selection step 
and a much more complicated clustering step (which combines several methods including PCA, k-nearest neighborhood 
algorithm, and modularity optimization).  Seurat needs $4$ tuning parameters: $m, N, k_0, \delta$, 
where $m$ is the number of selected features in the feature selection step, 
and $N, k_0, \delta$ are for the clustering step, corresponding to the PCA part, the k-nearest neighborhood algorithm part, and the modularity optimization part, respectively. 

Below is a high-level sketch for Seurat (see \cite{Satija}) for more detailed description).  
Input: un-normalized $n \times p$ data matrix $X$, number of clusters $K$, and tuning parameters $m, N, k_0, \delta$.  Output: 
predicted class label vectors $\hat{Y} = (\hat{Y}_1, \hat{Y}_2, \ldots, \hat{Y}_n)'$.  
\begin{itemize}
\item (IF-step).  Select the $m$ features that are mostly variable. Obtain the $n \times m$ post-selection data matrix. 
\item (Clustering-step). Normalize the post-selection data matrix and 
obtain the first $N$ left singular vectors.  For each pair of subjects, compute how many neighbors (for each subject, 
we only count the $k_0$ nearest neighbors) 
they share with each other, and use the results to construct a shared nearest neighborhood (SNN) graph. 
Cluster the class labels by applying a modularity optimization algorithm to the SNN graph, where we need 
a resolution parameter $\delta$. 
\end{itemize}
An apparent limitation of Seurat is that it needs $4$ tuning parameters. 
Following the recommendations by \cite{SeuratR}, we may take $(N, k_0) = (50, 20)$, but 
it remains unclear how to select $(m, \delta)$. 

SC3 was first presented by \cite{Vladimir}. To be consistent with many other methods 
we discuss in this paper, we may view SC3 as containing two main steps, a
gene filtering step and a clustering step. Similar to Seurat, 
the clustering step of SC3 is much more complicated than that of IF-PCA, 
where the main idea is to apply PCA many times (each for a different number 
of leading singular vectors) and use the results to construct a matrix of consensus.  
We then cluster all subjects into $K$ groups by applying the classical hierarchical clustering method to the consensus matrix. SC3 uses one tuning parameter $x_0$ in the gene filtering step, 
and two tuning parameters $d_0$ and $k_0$ in the clustering-step, corresponding to the 
PCA part and the hierarchical clustering part, respectively. 

Below is a high-level sketch for SC3 (see \cite{Vladimir}) for more detailed description).  
Input: un-normalized $n \times p$ data matrix $X$, true number of clusters $K$,  and tuning parameters $x_0, d_0, k_0$.  Output: predicted class label vectors $\hat{Y} = (\hat{Y}_1, \hat{Y}_2, \ldots, \hat{Y}_n)'$.  
\begin{itemize}
\item (Gene filtering-step). 
Removes genes/transcripts that are either expressed (expression value is more than 2) in less than $x_0 \%$ of cells   or expressed (expression value is more than 0) in at least $(100-x_0)\%$ of cells.  
 This step may reduce a significant fraction of features, and we consider it to be 
more like a feature selection step than a preprocessing step. 
\item (Clustering-step).  First, we take a log-transformation of the post-filtering data matrix and 
construct an $n \times n$ matrix $M$, where $M(i,j)$ is some kind of distances (e.g., Euclidean, Pearson, Spearman) 
between subject $i$ and $j$.  Second, Let $\widehat{H} = [\hat{\eta}_1, \ldots, \hat{\eta}_d]$, where 
$\hat{\eta}_k$ is the $k$-th singular vector of $M$ (or alternatively, of the normalized graph Laplacian matrix of $M$).   
Third, for $d = 1, 2, \ldots, d_0$, we cluster all $n$ subjects to $K$ classes by applying the k-means to the rows of the $n \times d$ sub-matrix of $\widehat{H}$ consisting of the first $d$ columns, and use the results to build a consensus matrix using the Cluster-based Similarity Partitioning Algorithm (CSPA) \citep{strehl2002cluster}. Finally, 
we cluster the subjects by applying the classical hierarchical clustering to the consensus matrix with 
$k_0$ levels of hierarchy. 
\end{itemize}
Following the recommendation by \cite{Vladimir}, we set $(x_0, d_0) = (6, 15)$ and take $k_0$ to be the true number of 
 clusters $K$. Such a tuning parameter choice may work effectively in some cases, but for more general cases, 
we may (as partially mentioned in \cite{Vladimir}) need more complicated tuning.

In summary, on a high level, we can view both Seurat and SC3 as two-stage 
algorithms, which consist of a feature selection step and a clustering step, just as in IF-PCA.  
However, these methods use more complicated clustering steps where the key is combining 
{\it many different clustering results to reach a consensus}; note that the Shared Nearest Neighborhood (SNN) 
in Seurat can be viewed a type of consensus matrix. 
Such additional miles taken in Seurat and SC3  may help reduce the clustering error rates, but also 
make the algorithms conceptually more complex, computationally more expensive,   and theoretically more difficult to analyze.

\subsection{A brief summary of all the methods} 
\label{subsec:method-sum} 
We have introduced about 10 different methods, 
some of which (e.g., IF-PCA(X),  IF-VAE, IF-VAE(X)) were never proposed before.  
Among these methods, VAE is a popular unsupervised deep learning approach, 
Seurat and SC3 are especially popular in clustering with single-cell data, 
and IF-PCA is a conceptually simple method which was shown to be effective 
in clustering with gene microarray data before. 
Note that some of the methods are conceptually similar to each other with some small differences (though 
it is unclear how different their empirical performances are). 
For example, many of these methods are two-stage methods, containing an IF-step and a clustering-step. In the 
IF-step, we usually use the normalized data matrix $W$. In the clustering-step, we may use either $W$ or the un-normalized data matrix $X$.  To summarize all these methods and especially to clarify the small differences between similar methods, 
we have prepared a table below; see Table \ref{tab:method-sum} for details. 
\begin{table}[h] 
\begin{center} 
\scalebox{0.75}{
\begin{tabular}{  |c |c |c| c| c | c| c|c |c|c|c| } 
 \hline
& PCA& SpecGem& VAE&VAE(X)&IF-PCA&IF-PCA(X)&IF-VAE&IF-VAE(X) & Seurat & SC3 \\
 \hline
IF-step& NA &NA&NA&NA&$W$&$W$&$W$&$W$ & $X$   & $X$  \\
\hline
Clustering-step&$X$ or $W$ &NA&$W$&$X$&$W$&$X$&$W$&$X$ & $X$ & $X$ \\
 \hline 
\end{tabular} }
\caption{A summary of all methods discussed in this section. This table clarifies the 
small differences between similar methods. Take the column IF-PCA(X) for example: 
``$W$" on row 2 means that the IF-step of this method is applied to the normalized data matrix $W$ defined in (\ref{normalization}), 
and ``$X$" on row 3 means the  clustering-step is applied to the un-normalized data matrix $X$ (NA: not applicable).}
\label{tab:method-sum}
\end{center}  
\end{table}

\section{Result} \label{sec:result}

Our study consists of two parts. In Section \ref{subsec:realdata1}, 
we compare  IF-VAE with several other methods using $10$ microarray data sets. 
In Section \ref{subsec:realdata2}, we compare  IF-VAE with several other methods, including the popular approaches of Seurat and SC3,  using $8$ single-cell data sets. In all these data sets, the class labels are given. However, we do not use the class labels in any of the clustering approaches; we only use them when we evaluate the error rates.  The code for numerical results in this section can be found at \url{https://github.com/ZhengTracyKe/IFPCA}. The 10 microarray data sets can be downloaded at \url{https://data.mendeley.com/datasets/cdsz2ddv3t}, and the 8 single-cell RNA-seq data sets can be downloaded at \url{https://data.mendeley.com/drafts/nv2x6kf5rd}.

\subsection{Comparison of clustering approaches with $10$ microarray data sets}
\label{subsec:realdata1} 

Table \ref{tab:mic-data} tabulates 10 gene microarray data sets (alphabetically) studied in \cite{JinWang}.  Here,  Data sets 1, 3, 4, 7, 8, and 9 were analyzed and cleaned in \cite{Dettling},  Data sets 2, 6, 10 were analyzed and grouped into two classes in \cite{Yousefi}, among which Data set 10 was cleaned by \cite{JinWang} in the same way as by \cite{Dettling}.  Data set 5 is from \cite{Gordon}. 

\begin{table}[tbh] 
\begin{center} 
\scalebox{.95}{
\begin{tabular}{  |c |l| l|| c| c | c| } 
 \hline
\#& Data Name& Source& $K$ & $n$ & $p$\\
 \hline
1&  Brain&Pomeroy (02)  &5&42 &5597\\
2 & Breast Cancer &Wang et al. (05)& 2&276   &22,215\\
3&Colon Cancer & Alon et al. (99)&  2&62&  2000\\
4&Leukemia& Golub et al. (99)&2&72& 3571\\
5&Lung Cancer(1)& Gordon et al. (02)& 2&181& 12,533\\ 
6&Lung Cancer(2)& Bhattacharjee et al. (01)&2&203& 12,600\\
7&Lymphoma& Alizadeh et al. (00)&3&62& 4026\\
8&Prostate Cancer& Singh et al. (02)&2&102& 6033\\
9&SRBCT& Kahn (01)&4&63& 2308\\
10&SuCancer& Su et al (01)&2&174 & 7909\\
 \hline
\end{tabular}}
\caption{The $10$ gene microarray data sets analyzed in Section \ref{subsec:realdata1}  ($n$: number of subjects; $p$: number of genes; $K$: number of clusters).}
\label{tab:mic-data}
\end{center}
\end{table}

First, we compare the IF-VAE approach introduced in Section \ref{subsec:IF-VAE} with four existing clustering methods: 
(1) the classical kmeans; (2) Spectral-GEM (SpecGem) \citep{SpecGem}, which is essentially classical PCA combined with a Laplacian normalization; (3) the orthodox IF-PCA \citep{JinWang}, which adds a feature selection step prior to spectral clustering (see Section \ref{subsec:IF-PCA} for details);  (4)  The VAE approach, which uses VAE for dimension reduction and then runs kmeans clustering (see Section \ref{subsec:VAE} for details).  
Among these methods, SpecGem and VAE involve dimension reduction, and IF-PCA and IF-VAE use both dimension reduction and feature selection. 
For IF-PCA, VAE and IF-VAE, we can implement the PCA step and the VAE step to either the original data matrix $X$ or the normalized data matrix $W$. The version of IF-PCA associated with $X$ is called IF-PCA(X), and the version associated with $W$ is still called IF-PCA; similar rules apply to VAE and IF-VAE. Counting these variants, we have a total of 8 different algorithms. 

Table~\ref{tab:mic-error} shows the numbers of clustering errors (i.e., number of incorrectly clustered samples, subject to a permutation of $K$ clusters) of these methods. The results of SpecGem and IF-PCA are copied from \cite{JinWang}. We implemented kmeans using the Python library \texttt{sklean}, wrote Matlab code for IF-PCA(X), and wrote Python code for the remaining four methods. 
The IF-step of IF-VAE needs no tuning. 
In the VAE-step of IF-VAE,  we fix the latent dimension as $d=25$ and use a traditional architecture in which both the encoder and decode have one hidden layer; the encoder uses the ReLU activation and the decode uses the sigmoid activation; when training the encoder and decoder, we use a mini-batch stochastic gradient descent with 50 batches, 100 epochs, and a  learning rate of 0.0005. The same neural network architecture and tuning parameters are applied to VAE.  
We note that the outputs of these methods may have randomness due to the initialization in the kmeans step or in the VAE step. 
For VAE, IF-VAE, and IF-VAE(X) we repeat the algorithm 10 times and report the average clustering error.  
For kmeans, we repeat it for 5 times (because the results are more stable); for IF-PCA(X), we repeat it 20 times. 
We use the clustering errors to rank all 8 methods for each data set; in the presence of ties, we assign ranks in a way such that the total rank sum is 36 (e.g., if two methods have the smallest error rate, we rank both of them as 1.5 and rank the second best method as 3; other cases are similar). 
The average rank of a method is a metric of its overall performance 
across multiple data sets. 
Besides ranks, we also compute {\it regrets}: 
For each data set, the {\it regret} of a method is defined to be $
r=(e - e_{min})/(e_{max}-e_{min})$,  
where $e$ is the clustering error of this method, and $e_{max}$ and $e_{min}$ are the respective maximum and minimum clustering error among all the methods. The average regret also measures the overall performance of a method (the smaller, the better).

\setlength{\tabcolsep}{2.5pt}
\begin{table}[tbh]
\begin{center}
\scalebox{0.9}{
\begin{tabular}{  | l | c c  c c c ccc|}
 \hline
Dataset& kmeans & SpecGem& IF-PCA & IF-PCA(X) & VAE & VAE(X) & IF-VAE & IF-VAE(X) \\
 \hline
Brain  &14  & 6 &  11 & 7 & 14 & 17 & 21& 21\\
Breast Cancer & 121&  121& 112& 91 &105&130 & 120 & 118\\
Colon Cancer  &28& 30& 25  &26 &  29&  23 & 25 & 25 \\
Leukemia &  2 &21&5  &3 &28 & 17 & 20  & 12\\
Lung Cancer(1)&18  & 22&  5&24& 21 & 64 &   6 & 7 \\ 
Lung Cancer(2)& 44 &88&  44 &45& 66 & 80 &  44 & 44\\
Lymphoma& 1  &14& 1  &18 & 23 & 22& 16  & 10 \\
Prostate Cancer &43  &43& 39& 44 & 41 &45 &42 &41\\
SRBCT& 28 &32& 28   &24&33 &26 & 30 & 23\\
SuCancer&83 & 85&58 & 57 & 62 & 60 & 57 & 57\\
\hline 
Rank(mean)& 4.3 &6.1& {\bf 2.65}  &3.9 & 5.7 &5.8& 4.3 & 3.25 \\
Rank(SD)&2.07& 2.20& 1.18 &2.33& 2.20& 2.35 &1.90 &1.74\\
Regret(mean)&0.43 &0.69& {\bf 0.18}& 0.26& 0.60& 0.65& 0.46& 0.31\\
Regret(SD)&0.35& 0.33 &0.22& 0.32 &0.33 &0.39& 0.36& 0.33\\
 \hline
\end{tabular}}
\caption{Comparison of clustering errors of different methods on the 10 microarray data
sets in Table \ref{tab:mic-data}. IF-PCA has the smallest average rank and average regret (boldface) and 
is regarded as the best on average.}
\label{tab:mic-error}
\end{center}
\end{table}
\setlength{\tabcolsep}{6pt}

There are several notable observations. First, somewhat surprisingly, the simple and tuning-free method, IF-PCA, has the best overall performance. It has the lowest average rank among all 8 methods and achieves the smallest number of clustering errors in 4 out of 10 data sets. We recall that the key idea of IF-PCA is to add a tuning-free feature selection step prior to dimension reduction. The results in Table~\ref{tab:mic-data} confirm that this idea is highly effective on microarray data and hard to surpass by other methods. 
Second, VAE (either on $W$ or on $X$), which combines k-means with nonlinear dimension reduction, significantly improves kmeans on some ``difficult" datasets, such as BreastCancer, ColonCancer and SuCancer. 
However, for those ``easy" data sets such as Leukemia and Lymphoma, VAE significantly underperforms kmeans. 
It suggests that the nonlinear dimension reduction is useful mainly on ``difficult" data sets. 
Third, IF-VAE (either on $W$ or on $X$) improves VAE in the majority of data sets. 
In some data sets such as LungCancer(1), the error rate of IF-VAE is much lower than that of VAE.  
This observation confirms that the IF step plays a key role in reducing the clustering errors. 
\cite{JinWang} made a similar observation by combining the IF step with linear dimension reduction by PCA.
Our results suggest that the IF step continues to be effective when it is combined with nonlinear dimension reduction by VAE. 
Last, IF-VAE(X) achieves the lowest error rate in 3 out of 10 data sets, and it has the second lowest average rank among all 8 methods. 
Compared with IF-PCA (the method with the lowest average rank), IF-VAE(X) has an advantage in 3 data sets (BreastCancer, SRBCT and SuCancer) but has a similar or worse performance in the other data sets. These two methods share the same IF step, hence, the results imply that the nonlinear dimension reduction by VAE has an advantage over the linear dimension reduction by PCA only on ``difficult" data sets. 

\begin{figure}[ht]
\begin{center}
\includegraphics[width=0.44\textwidth, trim=0 0 0 50, clip=true]{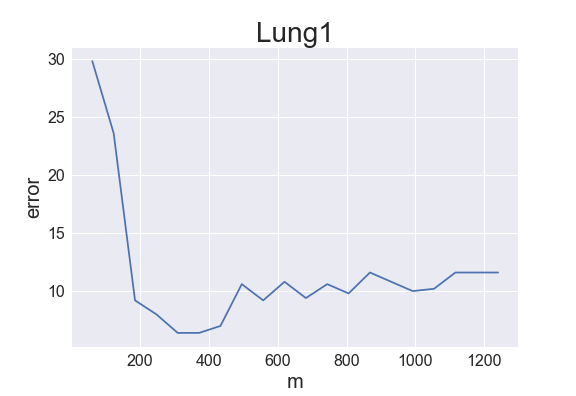}
\end{center}
\caption{Clustering errors of IF-VAE(X) as a function of the number of selected features in the IF step (data set: LungCancer(1); y-axis: number of clustering errors; x-axis: number of selected features).}\label{fig:lym}
\end{figure}

Next, we study IF-VAE(X) more carefully on the LungCancer(1) data set. Recall that the IF step ranks all the features using KS statistics and selects the number of features by a tuning-free procedure. 
We use the same feature ranking but manually change the number of retained features. For each $m$, we select the $m$ top-ranked features, perform VAE on the unnormalized data matrix $X$ restricted to these $m$ features, and report the average number of clustering errors over 5 repetitions of VAE. Figure~\ref{fig:lym} displays the number of clustering errors as a function of $m$.
An interesting observation is that as $m$ increases, the clustering error first decreases and then increases (for a good visualization, Figure~\ref{fig:lym} only shows the results for $m$ between 1 and $0.1p$; we also tried larger values of $m$ and found that the number of clustering errors continued to increase; especially, the number errors increased fast when $m>4000$). A possible explanation is as follows: when $m$ is too small, some influential features are missed, resulting in weak signals in the VAE step; when $m$ is too large, too many non-influential features are selected, resulting in large noise in the VAE step.  
There is a sweet spot between 200 and 400, and the tuning-free procedure in the IF step selects $m=251$. 
Figure~\ref{fig:lym} explains why IF step benefits the subsequent VAE step. 
A similar phenomenon was discovered in \cite{JinWang}, but it is for PCA instead of VAE.

{\it {\bf Remark 1} (Comparison with other clustering methods for microarray)}:  
\cite{JinWang} reported the clustering errors of several classical methods on these 10 microarray data sets. 
We only include kmeans and SpecGem in Table~\ref{tab:mic-error}, because kmeans is the most widely-used generic clustering methods and SpecGem is specially designed for microarray data. The table below shows the clustering errors of other methods reported in \cite{JinWang}, including kmeans++ (a variant of kmeans with a particular initlization) and hierarchical clustering. It suggests that these methods significantly underperform IF-PCA.

\begin{table}[h]
\begin{center}
\scalebox{.85}{
\begin{tabular}{  | l | c c  c   cc ccccc|}
 \hline
& Brain &Breast&Colon&Leuk&Lung1&Lung2&Lymph&Prostate&SRBCT&Su\\
 \hline
kmeans++&18&119&29&19&35&89&20&44&33&80\\
Hier&22&138&24&20&32&61&29&49&34&78\\
\hline
IF-PCA&11&112&25&5&5&44&1&39&28&58\\
 \hline
\end{tabular}}
\caption{The clustering errors of kmeans++ and hierarchical clustering on the 10 microarray data sets (the clustering errors of IF-PCA are listed for reference).}\label{tab:micro-other}
\end{center}
\end{table}

\subsection{Comparison of clustering approaches on $8$ single-cell RNA-seq data sets} \label{subsec:realdata2}

Table \ref{tab:rna-data} tabulates 8 single-cell RNA-seq data sets. 
The data were downloaded from the Hemberg Group at the Sanger Institute (\url{https://hemberg-lab.github.io/scRNA.seq.datasets}).   It contains scRNA-seq data sets from Human and Mouse. Among them, we selected 8 data sets that have a sample size between 100 and 2,000 and can be successfully downloaded and pre-processed using the code provided by Hemberg Group under the column `Scripts'. The data sets Camp1, Camp2, Darmanis, Li and Patel come from Human, and the data sets Deng, Goolam and Grun come from Mouse.  Each data matrix contains the log-counts of the RNA-seq reads of different genes (features) in different cells (samples). The cell types are used as the true cluster labels to evaluate the performances of clustering methods.   We first pre-processd all the data using the code provided by Hemberg Group, then features (genes) with fractions of non-zero entries $<5\%$ are filtered out.  
The resulting dimension for all data sets are shown in Table~\ref{tab:rna-data}. 
\begin{table}[h]
\begin{center}
\scalebox{0.95}{
\begin{tabular}{ |c |l || c | c |c|}
 \hline
 \#&Dataset&$K$ & $n$ & $p$ \\
 \hline
1&Camp1&7&777&13,111\\
2&Camp2&6&734&11,233\\
3&Darmanis&9 &466&13,400 \\
4&Deng  &6&268 &16,347\\
5&Goolam &5&124& 21,199\\
6&Grun&2&1502& 5,547\\
7&Li&9&561&25,369\\
8&Patel&5&430&5,948\\
 \hline
\end{tabular}}
\caption{Single-cell RNA-seq data sets investigated in this paper. ($n$: number of cells; $p$: number of genes; $K$: number of cell types)}
\label{tab:rna-data}
\end{center}
\end{table}

We compare IF-VAE with three other existing methods: (1) the orthodox IF-PCA \citep{JinWang}, (2) Seurat \citep{Satija} and (3) SC3 \citep{Vladimir}.  The orthodox IF-PCA was proposed for subject clustering on microarray data. It is the first time this method is applied to single-cell data. Seurat and SC3 are two popular methods clustering single-cell RNA-seq data (see Sections~\ref{subsec:SC3} for details). 
As discussed in Section~\ref{subsec:SC3}, Seurat and SC3 implicitly use some feature selection ideas and some dimension reduction ideas, but they are much more complicated than IF-PCA and have several tuning parameters. Seurat has 4 tuning parameters, where $m$ is the number of selected features,  $N$ is the number of principal components in use, $k_0$ is the number of clusters in k-nearest neighbors, and $\delta$ is a `resolution' parameter. 
We fix $(m, N, k_0)=(1000, 50, 20)$ for all data sets  (the values of $(N,k_0)$ are the default ones; the default value of $m$ is $2000$, but we found that $m$=$1000$ gives the same results on the 8 data sets and is faster to compute). We choose a separate value of $\delta$ for each data set in a way such that the resulting number of clusters from a modularity optimization is exactly $K$ (details can be found in \cite{waltman2013smart}).  
Seurat is implemented by the R package \href{https://satijalab.org/seurat/index.html}{\texttt{Serut}} \citep{SeuratR}. 
SC3 has 3 tuning parameters, where $x_0\%$ is a threshold of cell fraction used in the gene filtering step, $d_0$ is the number of eigenvectors in use, and $k_0$ is the level of hierarchy in the hierarchical clustering step. 
We fix $(x_0, d_0)=(10, 15)$ and set $k_0$ as the number of true clusters in each data set. SC3 is implemented using the R package \href{https://bioconductor.org/packages/release/bioc/html/SC3.html}{\texttt{SC3}} \citep{Vladimir}. We observed that SC3 output an NA value on the Patel data set, because the gene filtering step removed all of the genes. To resolve this issue, we introduced a variant of SC3 by skipping the gene filtering step. This variant is called SC3(NGF), where NGF stands for `no gene filtering.' Seurat, SC3 and SC3(NGF) can only be applied to the unnormalized data matrix $X$. 
These methods also have randomness in the output, but the standard deviation of clustering error is quite small; hence, we only run 1 repetition for each of them. The implementation of IF-PCA, IF-PCA(X), IF-VAE and IF-VAE(X) are the same as in Section~\ref{subsec:realdata1}.

\begin {table}[h]
\centering
\scalebox{0.9}{
\begin{tabular}{ |l || c  c cc c c c|}
  \hline
 Dataset & Seurat & SC3 &SC3(NGF) &IF-PCA&IF-PCA(X)&IF-VAE&IF-VAE(X)\\
  \hline
 Camp1&0.637& 0.750&0.627&0.738&0.736&  0.660  &0.700\\
Camp2&0.661& 0.713&0.759&0.601&0.656&  0.393  &0.491\\
Darmanis&0.682 & 0.826&0.867& 0.635& 0.747&  0.406  &0.617 \\
Deng &0.530&0.590 &0.754 & 0.791 &0.588&  0.607 & 0.687\\
Goolam &0.621& 0.758&0.629 & 0.637& 0.700&  0.612  &0.703\\
Grun&0.994& 0.509 &0.511&0.740& 0.657&0.595   &0.753\\
Li & 0.934&  0.938 &  0.980& 0.889& 0.968& 0.848 &0.853\\
Patel&0.898&NA &0.995& 0.795&0.934&  0.325  &0.465\\
\hline
Rank (mean)&3.5 &NA & {\bf 2.75} &  3.0& {\bf 2.75}& 5.38& 3.63\\
Rank (SD)&1.7 & NA&2.3 &  1.3& 1.2& 0.9& 1.6\\
Regret (mean)& 0.50& NA & 0.37&  0.40& {\bf 0.28}& 0.90& 0.53\\
Regret (SD) &0.4& NA & 0.5 & 0.3&0.3& 0.1& 0.3\\
 \hline
\end{tabular}
}
\caption {Comparison of the clustering accuracies with the 8 single-cell RNA-seq data
sets in Table \ref{tab:rna-data}. The result for SC3 on Patel is NA, because all genes are removed in the gene filtering step; for this reason, we exclude SC3 when calculating the rank and the regret. To resolve this issue, we also introduce a variant of SC3 by skipping the gene filtering step.  This variant is called 
SC3(NGF), where `NGF' stands for no gene filtering. It has a better performance than the original SC3. 
Note that IF-PCA(X) is regarded as the best on average:   it has the smallest average regret (boldface) and average rank (boldface). Note also that the standard deviation (SD) of its rank is only about $50\%$ of that of SC3(NGF).}
\label{tab:rna-error}
\end{table}

Table~\ref{tab:rna-error} contains the clustering accuracies (number of correctly clustered cells divided by the total number of cells) of different methods. 
For each data set, we rank all 6 methods (excluding SC3) by their clustering accuracies (the higher accuracy, the lower rank). SC3 is excluded in rank calculation, because it outputs NA on the Patel data set. Instead, we include SC3(NGF), a version of SC3 that resolves this issue on Patel and has better performances in most other data sets; this gives more favor to SC3 in the comparison. 
For each data set, we also compute the regret of each method (the same as in Section~\ref{subsec:realdata1}). Similarly, we exclude SC3 but include SC3(NGF) in the regret calculation. 
Each method has a rank and a regret on each data set. The last 4 rows of Table~\ref{tab:rna-error} show the mean and standard deviation of the 8 ranks of each method, as well as the mean and standard deviation of the 8 regrets of each method. 

We make a few comments. First, if we measure the overall performance on 8 data sets using the average rank, then IF-PCA(X) and SC3(NGF) are the best. 
If we use the average regret as the performance metric, then IF-PCA(X) is the best method.  
Second, a closer look at SC3(NGF) and IF-PCA(X) suggests that their performances have different patterns. 
SC3(NGF) is ranked 1 in some data sets (e.g., Camp2, Darmanis, etc.) but has low ranks in some other data sets (e.g., Goolam, Grun, etc.). 
In contrast, IF-PCA(X) is ranked 2 in almost all data sets. Consequently, IF-PCA(X) has a smaller rank standard deviation, even though the two methods have the same average rank. One possible explanation is that SC3 is a complicated method with several tuning parameters. For some data sets, the current tuning parameters are appropriate, and so SC3 can achieve an extremely good accuracy; for some other data sets, the current tuning parameters are probably inappropriate, resulting in an unsatisfactory performance. In comparison, IF-PCA is a simple and tuning-free method and has more stable performances across multiple data sets. 
Third, IF-VAE(X) is uniformly better than IF-VAE, hence, we recommend applying IF-VAE to the unnormalized data matrix instead of the normalized one.  
Last, IF-VAE(X) significantly improves IF-PCA(X) on Deng and Grun. This suggests that the nonlinear dimension reduction by VAE is potentially useful on these two data sets. In the other data sets, IF-VAE(X) either under-performs IF-PCA(X) or performs similarly. 

In terms of computational costs, Seurat is the fastest, and IF-PCA is the second fastest. VAE and SC3 are more time-consuming, where the main cost of VAE arises from training the neural network and the main cost of SC3 arises from computing the $n\times n$ similarity matrix among subjects. 
For a direct comparison, we report the running time of different methods on the Camp1 dataset ($n=777$ and $p=13111$). IF-PCA is implemented in Matlab and takes about 1.7 minutes. 
VAE and IF-VAE are implemented in Python, where the VAE steps are conducted using the Python library \texttt{keras}. The running time of VAE is 2.7 minutes, and the running time of IF-VAE is 1.4 minutes. 
SC3 is implemented via the package \texttt{SC3} of Bioconductor in R, and it takes 3 minutes. 
Seurat is implemented using the R package \texttt{Seurat} and takes only 6 seconds.

{\it {\bf Remark 2} (Using ARI as the performance metric)}: The adjusted rand index (ARI) is another commonly-used metric for clustering performance. In Table~\ref{tab:rna-ARI}, we report the ARI of different methods and recalculate the ranks and regrets. The results are quite similar to those in Table~\ref{tab:rna-error}.

\begin {table}[h]
\centering
\scalebox{0.85}{
\begin{tabular}{ |l || c   cc c c c c|}
  \hline
 Dataset &Seurat&SC3&SC3(NGF)&IF-PCA&IF-PCA(X)&IF-VAE&IF-VAE(X)\\
  \hline
Camp1&0.534& 0.768&0.526&0.628&0.627&  0.606  &0.615\\
Camp2&0.443&  0.577&0.502& 0.410&0.493&  0.162 &0.304\\
Darmanis&0.480 &0.682&0.784&  0.489& 0.650&  0.219 &0.525  \\
Deng &0.442&0.646&0.669& 0.771 &0.477&  0.487& 0.555  \\
Goolam  &0.543& 0.687&0.544&0.356& 0.562&  0.410  &0.534 \\
Grun& 0.969&-0.066&-0.060& 0.135& 0.102&0.023 &0.137\\
Li & 0.904&0.951&0.968&  0.797& 0.940& 0.798 &0.792\\
Patel &0.790& NA& 0.989& 0.598&0.850&  0.173 &0.235\\
\hline
Rank (mean)& 3.62 &NA&{\bf 2.50}& 3.50&{\bf 2.50}& 5.00 &3.88\\
Rank (SD)&1.60&NA& 2.20&  1.77& 1.31 &0.93& 1.36  \\
Regret (mean)& 0.42&NA& 0.30&   0.51 &{\bf 0.29}& 0.84 &0.59 \\
Regret (SD) & 0.37& NA&0.44&0.40 &0.37& 0.27 &0.33 \\
 \hline
\end{tabular}}
\caption {The values of adjusted rand index (ARI) for the same datasets and methods as in Table \ref{tab:rna-error}. Similar, the average rank and regret of SC3 is denoted as NA, for it generated NA on the Patel data set.}
\label{tab:rna-ARI}
\end{table}

{\it {\bf Remark 3} (Comparison with RaceID)}: Besides Seraut and SC3, there are many other clustering methods for single-cell data (e.g., see \cite{scRNA-algo} for a survey). RaceID \citep{RaceID} is a recent method. It runs an initial clustering, followed by an outlier identification; and the outlier identification is based on a background model of combined technical and biological variability in single-cell RNA-seq measurements. We now compare IF-PCA(X) and IF-VAE(X) with RaceID (we used the R package \href{https://cran.r-project.org/web/packages/RaceID/index.html}{\texttt{RaceID}} and set all tuning parameters to be the default values in this package). 
We observe that IF-PCA(X) and IF-VAE(X) outperform RaceID on most datasets. One possible reason is that the outlier identification step in RaceID is probably more suitable for applications with a large number of cells (e.g., tens of thousands of cells). 
\begin{table}[h]
\centering
\scalebox{.88}{
\begin{tabular}{  | l | c c  c   cc ccc|}
 \hline
& Camp1&Camp2&Darmanis&Deng&Goolam&Grun&Li&Patel\\
 \hline
IF-PCA(X)&0.736&0.656&0.747&0.588&0.700&0.657&0.968&0.934\\
IF-VAE(X)&0.700&0.491&0.617&0.687&0.703&0.753&0.853&0.465\\
RaceID&0.645&0.425&0.290&0.630&0.443&0.583&0.624&0.542\\
 \hline
\end{tabular}}
\caption{Comparison of the clustering accuracies of IF-PCA(X), IF-VAE(X) and RaceID.}
\end{table}

{\it {\bf Remark 4} (Combining the IF-step with Seurat and SC3)}: 
We investigate if the IF-step of IF-PCA can be used to conduct feature selection for other clustering methods. To this end, we introduce IF-Seurat and IF-SC3(NGF), in which Seurat and SC3(NGF) are applied respectively to the post-selection unnormalized data matrix from the IF-step of IF-PCA. Table~\ref{tab:rna-IFplus} compares these two methods with their original versions. For Seurat, the IF-step improves the clustering accuracies on Camp1, Darmanis, and Patel, yields similar performances on Deng, Goolam Grun, and Li, and deteriorates the performances significantly on Camp2. 
For SC3, the IF-step sometimes yields a significant improvement (e.g., Camp1) and sometimes a significant deterioration (e.g., Deng). It is an interesting theoretical question when the current IF-step is suitable to combine with clustering methods other than PCA. 
\begin{table}[h]
\centering
\scalebox{.88}{
\begin{tabular}{  | l | c c  c   cc ccc|}
 \hline
& Camp1&Camp2&Darmanis&Deng&Goolam&Grun&Li&Patel\\
 \hline
Seurat&0.637&0.661&0.682&0.530&0.621&0.994&0.934&0.898\\
IF-Seurat & 0.647 & 0.485 & 0.779 & 0.526 & 0.597 & 0.986 & 0.879 & 0.937\\
\hline
SC3(NGF)&0.627&0.759&0.867&0.754&0.629&0.511&0.980&0.995\\
IF-SC3(NGF) & 0.724 & 0.702 & 0.796 & 0.489 & 0.637 & 0.550 & 0.998 & 0.981\\
 \hline
\end{tabular}}
\caption{Combinations of IF-Seurat with Seurat and IF-SC3(NGF) with SC3(NGF).}\label{tab:rna-IFplus}
\end{table}

\section{Phase transition for PCA and IF-PCA}
\label{sec:theory} 
\setcounter{equation}{0}   
Compared with VAE, Seurat, and SC3, an advantage of IF-PCA is that 
it is conceptually much simpler and thus comparably easier to analyze.  
In this section, we present some theoretical results and show that 
IF-PCA is optimal in a Rare/Weak signal setting.

We are interested in several intertwined questions.  
\begin{itemize} 
\item When the IF-step of the IF-PCA is really necessary.  As 
IF-PCA reduces to classical PCA when we omit the 
IF-step,  an equivalent question is when IF-PCA really has an advantage over 
of PCA. 
\item When IF-PCA is optimal in a minimax decision framework. 
\end{itemize}

To facilitate the analysis, we consider a high-dimensional clustering setting 
where $K = 2$ so we only have two classes.  We assume the two classes are equally likely so the class labels satisfy 
\begin{equation} \label{RW1a} 
Y_i \stackrel{iid}{\sim} 2 \mathrm{Bernoulli}(1/2) -1, \qquad 1\leq i\leq n;  
\end{equation} 
extension to the case where we replace the Bernoulli parameter $1/2$ by a $\delta \in (0,1)$ is comparably straightforward. 
We also assume that  the $p$-dimensional data vectors $X_i$'s are standardized, so that for a  contrast mean vector $\mu \in R^p$  ($I_p$ standards for the $p \times p$ identity matrix), 
\begin{equation} \label{RW1b} 
X_i =  Y_i  \mu + Z_i,   \qquad Z_i \stackrel{iid}{\sim} N(0, I_p), \qquad 1 \leq i \leq n. 
\end{equation}  
As before, write $Y = (Y_1, Y_2, \ldots, Y_n)'$,  $X = [X_1, X_2, \ldots, X_n]'  = [x_1, x_2, \ldots, x_p]$.
 It follows 
\[
X = Y \mu' + Z,  \qquad \mbox{where similarly $Z = [Z_1, Z_2, \ldots, Z_n]'  = [z_1, z_2, \ldots, z_p]$}. 
\] 
For any $1 \leq j \leq p$, we call feature $j$  an ``influential feature" or ``useless feature"  if $\mu(j) \neq 0$ 
and  a ``noise" or ``useless feature" otherwise. 
We adopt a  Rare/Weak model setting where ($\nu_a$ stands for point mass at $a$) 
\begin{equation} \label{RW1c} 
\mu(j) \stackrel{iid}{\sim} (1 - \eps_p) \nu_0 + (\eps_p/2)  \nu_{\tau_p} + (\eps_p/2) \nu_{-\tau_p}. 
\end{equation} 

For fixed parameters $0 < \theta, \beta, \alpha < 1$, 
\begin{equation} \label{RW1d} 
n =  n_p = p^{\theta}, \qquad \eps_p = p^{-\beta}, \qquad \tau_p  =  p^{-\alpha}. 
\end{equation} 
From time to time, we drop the subscript of $n_p$ and write $n = n_p$.   
For later use, let 
\begin{equation} \label{DefineS} 
s_p = p \eps_p \qquad \mbox{and} \qquad \mbox{$S_p(\mu) = \{1 \leq j \leq p: \mu(j) \neq 0\}$ be the support of $\mu$}. 
\end{equation} 
It is seen $|S_p(\mu)| \sim \mathrm{Bernoulli}(p, \eps_p)$ and $|S_p(\mu)| / s_p \sim 1$.  
Model (\ref{RW1a})-(\ref{RW1d})  models a scenario where $1 \ll n \ll p$ and 
\begin{itemize}  
\item (Signals are Sparse/Rare). The fraction of influential feature is $p^{-\beta}$, which $\goto 0$ rapidly as $p \goto \infty$,  
\item (Signals are individually Weak). The signal strength of each influential feature may be much smaller than $n^{-1/4}$ and  the signals are individually 
weak; it is non-trivial to separate the useful features from the useless ones.  
\item (No free lunch). Summing over $X$ either across rows (samples) or across columns (feature) would not 
provide any useful information for clustering decisions.
\end{itemize} 
The model is frequently used if we want to 
study the fundamental limits and phase transition associated with a high-dimensional statistical decision problem (e.g., 
classification, clustering, global testing). Despite the seeming simplicity, the RW model is 
actually very delicate to study, for it models a setting where the signals (i.e., useful features) 
are both rare and weak. See \cite{DJ04, DJ15, Jager, HJ, XieJC, ACV} for example. 
 
Compared with the model in \citep{JinWang} (which only considers one-sided signals, where 
all nonzero $\mu(j)$ are positive), our model allows two-sided signal and so is different. In particular, 
in our model, summing over $X$ either across rows or columns  would not 
provide any useful information for clustering decisions. 
As a result, the phase transition we derive below is different from those in \citep{JinWang}.

Consider a clustering procedure and let $\hat{Y} \in \mathbb{R}^n$ be the predicted class label vector. 
Note that for any $1 \leq i \leq n$, both $Y_i $ (true class label) and $\hat{Y}_i$ take values from $\{-1, 1\}$.  
Let $\Pi$ be the set of all possible permutations on $\{-1,1\}$. We measure the performance of $\hat{Y}$ by the  Hamming error rate:  
\begin{equation} \label{hamm1} 
\hamm_p(\hat{Y}, Y) = \hamm_p(\hat{Y}, Y;  \beta,\theta) =   n^{-1}  \inf_{\pi_0  \in \Pi}   \biggl\{ \sum_{i = 1}^n  P(\hat{Y}_i  \neq \pi_0  Y_i) \biggr\},    
\end{equation} 
where the probability measure is with respect to the randomness of $(\mu, Y, Z)$.

\subsection{A slightly simplified version of PCA and IF-PCA}  
\label{subsec:simplePCA} 
To facilitate analysis for Model (\ref{RW1a})-(\ref{RW1d}), we consider a slightly more idealized version of PCA and IF-PCA, 
where the main changes are (a) we skip the normalization step (as we assume the model is for data that is already normalized), and (b) we replace feature selection by  Kolmogorov-Smirnov statistics in IF-PCA
by feature selection by the $\chi^2$ statistics,  (c) we remove Efron's correction in IF-PCA (Efron's correction is especially useful for analyzing gene microarray data, but is not necessary for the current model), and (d) 
we skip the Higher Criticism Threshold (HCT) choice (the study on HCT is quite relevant for our model, but technically it 
is very long so we skip it).  
Note also the rank of the signal matrix $Y \mu'$ is $1$ in Model (\ref{RW1a})-(\ref{RW1d}), 
so in both PCA and the clustering step of IF-PCA, we should apply kmeans clustering to the  first singular vector of $X$ only.  
Despite these simplifications,  the essences of original PCA and IF-PCA are retained.  See below for 
more detailed description of the (simplified) PCA and IF-PCA.

In detail, to use PCA for Model (\ref{RW1a})-(\ref{RW1d}), we run the following.  
\begin{itemize} 
\item Obtain the first singular vector of $X$ and denote it by $\xi$ (this is simpler than $\hat{\xi}$; we are misusing the notation a little bit here).  
\item Cluster by letting $\hat{Y}_i = \mathrm{sgn}(\xi_i)$, $1 \leq i \leq n$. 
\end{itemize} 
To differentiable from PCA in Section \ref{subsec:PCA}, we may call the approach {\it the slightly simplified PCA}. 

Also, to use IF-PCA for Model (\ref{RW1a})-(\ref{RW1d}), we introduce the 
normalized $\chi^2$-testing scores for feature $j$ by 
\begin{equation} \label{Definepsi} 
\psi_j = (\|x_j\|^2 - n) / \sqrt{2n}.  
\end{equation} 
By elementary statistics,  
\[
\psi_j \sim  \left\{ 
\begin{array}{ll} 
N(\sqrt{(n/2)} \tau_p^2, 1), &\qquad \mbox{if feature $j$ is useful}, \\
N(0,1), &\qquad \mbox{otherwise}.   
\end{array}
\right. 
\] 
Fix a threshold 
\[
t_p^* = \sqrt{2 \log(p)}. 
\] 
The IF-PCA runs as follows. 
\begin{itemize}
\item (IF-step). Select feature $j$ if and only if $\psi_j \geq t_p^*$.  
\item (Clustering-step).  Let 
\[
\hat{S} = \{1 \leq j \leq p: \psi_j \geq t_p^*\}, 
\]
and let $X_{\hat{S}}$ be the post-selection data matrix (which is a sub-matrix of $X$ consisting of columns in $\hat{S}$). 
Let $\xi^* \in \mathbb{R}^n$ be the first singular vector of $\hat{X}_S$.  We cluster by letting 
\[
\hat{Y}_i = \mathrm{sgn}(\xi_i^*), \qquad 1 \leq i \leq p.  
\] 
\end{itemize} 
Similarly, to differentiate from the IF-PCA in Section \ref{subsec:IF-PCA}, we call this {\it the slightly simplified IF-PCA}.

\subsection{The computational lower bound (CLB)} 
We first discuss the computational lower bound (CLB).  
The notion of CLB is an extension of the classical information lower bound (LB) (e.g.,  the Cramer-Rao lower bound), and in comparison,  
\begin{itemize} 
\item Classical information lower bound  usually claims a certain goal is not achievable for any methods (which includes methods that are computationally NP hard). 
\item Computational lower bound usually claims a certain goal is not achievable for any methods with {\it a polynomial computational time}. 
\end{itemize} 
From a computational perspective, we highly prefer to have algorithms  with  a polynomial computation time. 
Therefore, compared with classical information lower bound, CLB is practically more relevant. 

Let $s_p = p \eps_p$. Note that in our model, the number of signals is 
$\mathrm{Bernoulli}(p, \eps_p)$, which concentrates at $s_p$. 
Recall that in our calibrations, $n  = p^{\theta}$ and $s_p = p^{1 - \beta}$, and the strength of individual signals is $\tau_p$. 
Introduce the critical signal strength by 
\[
\tau_p^*  =  \left\{ 
\begin{array}{ll} 
[p /(n s_p^2)]^{1/4}, &\qquad  \mbox{if $\beta < 1/2$  (so $s_p \gg \sqrt{p}$)}, \\  
n^{-1/4},          &\qquad \mbox{if $1/2 < \beta < (1 - \theta/2)$  (so $\sqrt{n} \ll s_p \ll \sqrt{p}$)}.  \\ 
s_p^{-1/2},   &\qquad \mbox{if $(1-\theta/2)  < \beta < 1$  (so $1 \ll s_p \ll \sqrt{n}$)}.  \\ 
\end{array} 
\right. 
\] 
We have the following theorem. 
\begin{thm} \label{thm:CLB} 
({\it Computational Lower Bound)}). 
Fix $(\theta, \beta) \in (0,1)^2$ and consider the clustering problem for Models (\ref{RW1a})-(\ref{RW1d}).    
As $p \goto \infty$, if $\tau_p / \tau_p^* \goto 0$, then 
for any clustering procedure $\hell$ with a polynomial computational time, 
  $\hamm_p(\hell,Y) \geq (1/2 + o(1))$. 
\end{thm} 
In other words, any ``computable clustering procedures" (meaning those with a polynomial computational time) fails in this case, where the error rate is approximately the same as that of random guess.  
The proof of Theorem \ref{thm:CLB} is long but is similar to that of \cite[Theorem 1.1]{JinKeWang}, 
so we omit it.    

Next, we study the performance of classical PCA and IF-PCA. But before we do that, 
we present a lemma on classical PCA in Section \ref{subsec:lemma}.   We state the lemma in a setting that is more general than Model (\ref{RW1a})-(\ref{RW1d}), but we will come back to Model (\ref{RW1a})-(\ref{RW1d})
in Section (\ref{subsec:theoryPCA}).

\subsection{A useful lemma on classical PCA} 
\label{subsec:lemma} 
Suppose we have a data matrix $X \in \mathbb{R}^{N, m}$ in the form of 
\begin{equation} \label{PCAmodel} 
X = Y \mu' + Z, \qquad  Y \in \mathbb{R}^N,  \mu \in  \mathbb{R}^m.  
\end{equation} 
In such a setting, we investigate when the PCA approach in Section \ref{subsec:simplePCA}  is successful. 
Recall that $\xi$ is the first singular vector of $X$. 
By basic algebra, 
it is the first eigenvector of the $N \times N$ matrix $XX'$, 
or equivalently, the first eigenvector of $XX' - m I_N$. 
Write 
\begin{align*} 
XX' - m I_N  & = \|\mu\|^2  YY' +   (Z Z'  - m I_N)   + (Y \mu' Z'    +   Z \mu Y')  \\
&  =   \|\mu\|^2 \cdot  YY'  +   (ZZ' - m I_N)  +    \mbox{secondary term}. 
\end{align*} 
In order for the PCA approach to be successful, we need that the spectral norm of  $\|\mu\|^2 YY'$ is much larger than that of $(ZZ' - m I_N)$. 
Note that $\|\mu\|^2 YY'$ is a rank-$1$ matrix where the spectral norm is $N \|\mu\|^2$. Also, by Random Matrix Theory \citep{vershynin2010introduction},   the spectral norm of 
$(ZZ' - m I_N)$ concentrates at  $(\sqrt{N} + \sqrt{m})^2 - m = N + 2 \sqrt{N m}$.  
Therefore, the main condition we need for the PCA approach to be successful is 
\begin{equation} \label{PCAmaincondition}  
N \|\mu\|^2 /  (N + 2 \sqrt{N m}) \goto \infty. 
\end{equation} 
We have the following lemma. 
\begin{lemma}  \label{lemma:PCA} 
Consider Model (\ref{PCAmodel}) where condition (\ref{PCAmaincondition}) holds and that $\|\mu\|^2 \gg \log(N + m)$.   
Let $\xi$ be the first left singular vector of $X$.  
When $\min\{N, m\} \goto \infty$,  with probability $1 - o(m^{-3})$, 
\[
\min\{ \| \sqrt{N} \xi + Y \|_{\infty},  \|\sqrt{N} \xi - Y \|_{\infty} \}   = o(1).  
\]   
\end{lemma} 

Lemma~\ref{lemma:PCA} is proved in the supplementary material. This result connects to the recent interests of studying entry-wise large-deviation bounds of eigenvectors \citep{abbe2020entrywise,fan2022asymptotic}. Our proof is based on a form of Taylor expansion of eigenvectors. Please see the supplementary material for details.

By Lemma \ref{lemma:PCA}, there is an error vector $r$ with $\|r\|_{\infty} = o(1)$ such that 
\[
\sqrt{N} \xi = \pm Y + r; \qquad \mbox{recall that $Y_i \in \{-1, 1\}$}. 
\] 
Therefore, if we let $\hat{Y}_i = \mathrm{sgn}(\xi_i)$ as in PCA approach in Section \ref{subsec:simplePCA}, then  except for a small probability,  
\[
\hat{Y} = \pm Y. 
\] 
This says that the PCA approach is able to fully recover the true class labels.

\subsection{Achievability of classical PCA and IF-PCA} 
\label{subsec:theoryPCA}  
We now come back to Model (\ref{RW1a})-(\ref{RW1d}) and study the behavior of classical PCA and IF-PCA in our setting. 
The computational limits of clustering has received extensive interests (e.g., \citep{luo2022tensor}). 
By the computational lower bound \citep{JinKeWang}, successful clustering by a computable algorithm 
is impossible when $\frac{\tau_p}{\tau_p^*} \goto 0$, so the interesting parameter range for PCA and 
IF-PCA is when 
\[
\tau_p/ \tau_p^* \goto \infty. 
\] 
We first discuss when feature selection by $\chi^2$-test is feasible. 
As before, let 
\[
\psi_j = (2n)^{-1/2} (\|x_j\|^2 - n)
\]
be the feature-wise $\chi^2$-testing scores, and recall that approximately,  
\[
\psi_j  \sim  \left\{ 
\begin{array}{ll} 
N(\sqrt{(n/2)} \tau_p^2, 1), &\qquad \mbox{if feature $j$ is useful}, \\
N(0,1), &\qquad \mbox{otherwise}.   
\end{array}
\right. 
\] 
We can view $\sqrt{(n/2)} \tau_p^2$ as the Signal-to-Noise ratio (SNR) for the 
$\chi^2$-test for a useful feature.   
We have two cases. 
\begin{itemize} 
\item (Less sparse case of $\beta < 1/2$). In this case, the number of useful features $s_p$ is much larger than $\sqrt{p}$ and $\tau_p^* \ll n^{-1/4}$,    
and the SNR of $\psi_j$ for a useful feature $j$ may be much smaller than $1$ even though $\tau_p / \tau_p^* \goto \infty$. 
In such a case, feature selection by the $\chi^2$-test is not useful.  
Consequently, except for a negligible probability,  the IF-step of IF-PCA selects all features, so IF-PCA reduces to PCA. 
\item (More sparse case of $\beta > 1/2$).  
 In this case, the number of useful features $s_p$ is much smaller than $\sqrt{p}$ and $\tau_p^* \geq n^{-1/4}$. 
  If $\tau_p / \tau_p^* \goto \infty$, then the 
SNR of $\psi_j$ $\goto \infty$ if $j$ is a useful feature. In such a case, feature selection maybe successful and IF-PCA is significantly different from PCA. 
\end{itemize} 

Consider the first case and suppose we apply the PCA approach in Section \ref{subsec:simplePCA} directly to matrix $X$. 
Applying Lemma \ref{lemma:PCA} with $(N, m) = (n, p)$ and noting that in this setting, 
\[
n \|\mu\|^2 \sim  n s_p  \tau_p^2, \qquad N +  2\sqrt{Nm} = p + 2 \sqrt{n p}  \sim 2 \sqrt{n p} \;\;  (\mbox{since $n \ll p$}), 
\] 
the PCA approach is successful if 
\[ 
n s_p \tau_p^2 / \sqrt{np} \goto \infty. 
\] 
Comparing this with the definition of $\tau_p^*$, this is equivalent to 
\[
\tau_p / \tau_p^* \goto \infty, \qquad \mbox{as $0 < \beta < 1/2$ in the current case}. 
\] 
We have the following theorem. 
\begin{thm} \label{thm:PCA} 
({\it Possibility Region for PCA}). 
Fix $(\theta, \beta) \in (0,1)^2$ and consider the clustering problem for Models (\ref{RW1a})-(\ref{RW1d}). 
Let $\hat{Y}^{pca}$ be the predicted class label vector by the PCA algorithm in Section \ref{subsec:simplePCA}.  
As $p \goto \infty$,  if 
\begin{equation} 
0 < \beta < 1/2  \; (\mbox{so $s_p / \sqrt{p} \goto \infty$}) \qquad \mbox{and} \qquad \frac{\tau_p}{\tau_p^*} \goto \infty,  
\end{equation} 
then $\hamm_p(\hell^{pca}, Y)  \goto 0$. 
\end{thm}    
 
Consider the second case, where we may have successful feature selection so it is desirable to use IF-PCA.
We assume
\begin{equation} \label{IF-PCA-condition1}  
\tau_p / \tau_p^* \geq (4 \log(p))^{1/4}, 
\end{equation} 
which is slightly stronger than that of $\tau_p^* / \tau_p  \goto \infty$.  
By the definition of $\tau_p^*$, we have that in the current case (where $1/2 < \beta < 1$) 
\begin{equation} \label{IF-PCA-condition1-Add} 
\tau_p^* \geq n^{-1/4}. 
\end{equation} 
 Recall that $S(\mu)$ is the true support of $\mu$ and 
\[
\hat{S} = \{1 \leq j \leq p: \psi_j \geq \sqrt{2 \log(p)}\}
\]
is the set of selected features in the IF-step of IF-PCA. Recall that 
\[
\psi_j  \sim  \left\{ 
\begin{array}{ll} 
N(\sqrt{(n/2)} \tau_p^2, 1), &\qquad \mbox{if feature $j$ is useful}, \\
N(0,1), &\qquad \mbox{otherwise}.   
\end{array}
\right. 
\] 
By (\ref{IF-PCA-condition1})-(\ref{IF-PCA-condition1-Add}), for any useful feature $j$, the SNR is 
\[
\sim \sqrt{(n/2)} \tau_p^2 \geq \sqrt{(n/2)} \sqrt{4 \log(p)} n^{-1/2} = \sqrt{2 \log(p)}. 
\]  
By elementary statistics, we have that approximately, 
\[
P(\hat{S} \neq  S)  = o(1),  \qquad \mbox{where for short $S = S(\mu)$; same below}. 
\]  
Therefore, except for a negligible probability,  
\[
X_{\hat{S}} = X_{S} = Y   \mu_S' + Z_{S},   
\] 
where similar as before,  $\mu_S$ is the sub-vector of $\mu$ with all entries restricted to $S$, and 
$X_{S}$ and $Z_S$ the sub-matrix of $X$ and $Z$ respectively, with columns restricted to $S$. 
Therefore,  in the clustering-step of IF-PCA, we are in effect applying the PCA approach of Section \ref{subsec:simplePCA} 
to $X_S$, where we recall $|S| / s_p \approx 1$.  
Applying Lemma \ref{lemma:PCA} with $(N, m) = (n, |S|)$ and noting that  
\[
n \|\mu_S\|^2 \sim  n  s_p  \tau_p^2, \qquad N +  2\sqrt{Nm} = n + 2 \sqrt{n |S|} \sim n + 2 \sqrt{n s_p},     
\] 
it follows that in order for the clustering-step of IF-PCA to be successful, we need 
\begin{equation} \label{IF-PCA-condition2} 
n s_p \tau_p^2 / (n + 2 \sqrt{n s_p}) \goto \infty, \qquad (\mbox{note that when $s_p \ll n$, this is equivalent to $s_p \tau_p^2 \goto \infty$}). 
\end{equation} 
Combining this with (\ref{IF-PCA-condition2}) and recalling that in the current case, 
$s_p \ll \sqrt{p}$,   IF-PCA is successful when  
\begin{equation} \label{IF-PCA-condition3} 
\left\{ 
\begin{array}{ll} 
\tau_p^2 \geq 2 \sqrt{\log(p) / n}, &\qquad \mbox{if $\sqrt{n} \ll s_p \ll \sqrt{p}$}, \\
s_p \tau_p^2  \goto \infty,    &\qquad \mbox{if $s_p \ll \sqrt{n}$}. \\
\end{array} 
\right. 
\end{equation} 
Comparing this with the definition of $\tau_p^*$, (\ref{IF-PCA-condition3}) hold if we assume 
\[
\tau_p/(\sqrt{\log(p)}  \tau_p^*)   \goto \infty, 
\] 
which is slightly stronger than that of $\tau_p / \tau_p^* \goto \infty$. 
We have the following theorem. 
\begin{thm} \label{thm:IF-PCA} 
({\it Possibility Region for IF-PCA}). 
Fix $(\theta, \beta) \in (0,1)^2$ and consider the clustering problem for Models (\ref{RW1a})-(\ref{RW1d}). 
Let $\hat{Y}^{ifpca}$ be the predicted class label vector by the PCA algorithm in Section \ref{sec:theory}.  
As $p \goto \infty$,  if 
\begin{equation} 
1/2 < \beta < 1  \; (\mbox{so $s_p / \sqrt{p} \goto 0$}) \qquad \mbox{and} \qquad \frac{\tau_p}{\sqrt{\log(p)} \tau_p^*}  \goto \infty, 
\end{equation} 
then in the IF-step of IF-PCA, 
\[
P(\hat{S} \neq S(\mu)) = o(1). 
\] 
Moreover, $\hamm_p(\hell^{ifpca}, Y)  \goto 0$. 
\end{thm}

\subsection{Phase transition} 
\begin{figure}[ht]
\begin{center}
\includegraphics[width=0.44\textwidth, trim=0 00 0 0, clip=true]{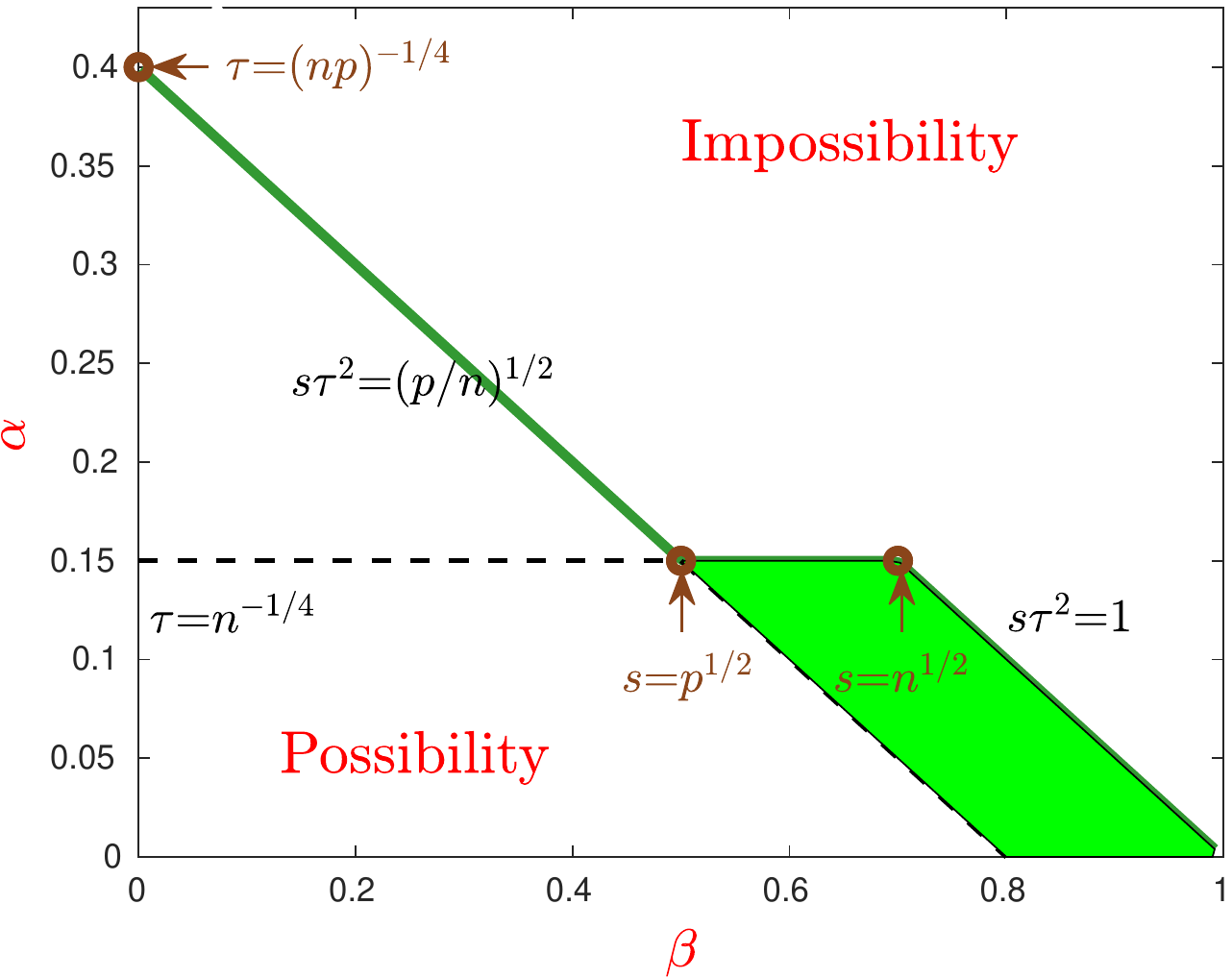}
\end{center}
\caption{ Phase transition for PCA and IF-PCA ($\theta = 0.6$).   The (three-segment) solid green line is 
$\alpha = \alpha^*(\beta, \theta)$,  which separates the whole region into the Region of Impossibility (top) 
and Region of Possibility (bottom). In the part of Region of Possibility ($\beta < 1/2$), feature selection is infeasible,   PCA is optimal, and IF-PCA 
reduces to PCA with an appropriate threshold.  In the right part ($\beta > 1/2$), it is desirable to conduct feature selection, and IF-PCA is optimal. However, PCA is non-optimal for parameters in the shaded green region.}\label{fig:Phase}
\end{figure}

Recall that $s_p  = p \eps_p$ and that in Model (\ref{RW1a})-(\ref{RW1d}),   
\[
n = n_p = p^{\theta}, \qquad \eps_p = p^{-\beta}, \qquad \tau_p = p^{-\alpha}.  
\] 
It follows 
\[
\tau_p^* = p^{- \alpha^*(\beta, \theta)},  \qquad \mbox{where}  \qquad 
\alpha^*(\beta, \theta) = 
\left\{
\begin{array}{ll}
(1 + \theta - 2 \beta)/4, &\qquad \mbox{if $0 < \beta < 1/2$}, \\
\theta/4, &\qquad  \mbox{if $1/2 < \beta < 1 - \theta/2$}, \\ 
(1-\beta)/2, &\qquad  \mbox{if $(1 - \theta/2) < \beta < 1$}.  \\
\end{array} 
\right.   
\] 
Fixing $0 < \theta < 1$, and consider the two-dimensional space where the two axes are $\beta$ and $\alpha$, respectively. 
Combining Theorem \ref{thm:PCA}-\ref{thm:IF-PCA}, the curve $\alpha = \alpha^*(\beta, \theta)$ partitions the region $\{(\alpha, \beta): 0 < \beta < 1, \alpha>0\}$ into two regions. 
\begin{itemize} 
\item {\it Region of Impossibility $\{(\alpha, \beta): \alpha >    \alpha^*(\beta, \theta), 0 < \beta <1\}$}. In this region, 
the Hamming clustering error rate of any methods with polynomial computation time is bounded away from $0$.  
\item {\it Region of Possibility $\{(\alpha, \beta): \alpha <  \alpha^*(\beta, \theta), 0 < \beta <1\}$}. The region further partitions into two parts:  $\beta < 1/2$ (left) and $\beta > 1/2$ (right).   
\begin{itemize} 
\item The left is the {\it less sparse case}  where the number of useful features $s_p \gg \sqrt{p}$. 
For any fixed $(\alpha, \beta)$  in this region,  the Hamming error rates of PCA are $o(1)$, so PCA  achieves the optimal phase transition. Also, in this case,  
the signals are too weak individually and feature selection is infeasible. Therefore, in the IF-step, 
the best we can do is to select all features, so IF-PCA reduces to PCA.  
\item The right is the {\it more sparse case}, where  the number useful features $s_p \ll  \sqrt{p}$. 
For any fixed $(\alpha, \beta)$  in this region, the Hamming error rates IF-PCA is $o(1)$, 
 so IF-PCA achieves the optimal phase transition.  Also in this case, the signals are strong enough individually and feature selection is desirable. Therefore, IF-PCA and PCA are significantly different.  
\item In particular, for any fixed parameters in the region $\{1/2 < \beta < 1,  (1 - \theta - 2 \beta) <  \alpha < (1 - \beta)/2 \}$ (shaded green region of Figure \ref{fig:Phase}),    the Hamming clustering error rate of IF-PCA is $o(1)$ but that of PCA is bounded away from $0$.    Therefore, PCA is non-optimal this particular region. 
\end{itemize} 
\end{itemize} 
See Figure \ref{fig:Phase} for details.

\section{Discussions}

IF-PCA is a simple and tuning-free approach to unsupervised clustering of high-dimensional data. The main idea of IF-PCA is a proper combination of the feature selection and the dimension reduction by PCA. 
In this paper, we make several contributions. First, we extend IF-PCA to IF-VAE, by replacing PCA with the variational auto-encoder (VAE), a popular unsupervised deep learning algorithm. 
Second, we study the theoretical properties of IF-PCA in a simple clustering model and derive the phase transitions. Our results reveal how the feature sparsity and the feature strength affect the performance of IF-PCA, and explain why IF-PCA can significantly improve the classical PCA. Third, we investigate the performances of IF-PCA and IF-VAE on two applications, the subject clustering with gene microarray data and the cell clustering with single-cell RNA-seq data, and compare them with some other popular methods. 

We discover that IF-PCA performs quite well in the aforementioned applications. Its success on microarray data was reported in \cite{JinWang}, but it has never been applied to single-cell data. To use IF-PCA on single-cell data, we recommend a mild modification of the original procedure called IF-PCA(X), which performs the PCA step on the unnormalized data matrix $X$ instead of the normalized data matrix $W$. On the 8 single-cell RNA-seq data sets considered in this paper, IF-PCA(X) has the second best accuracy in almost all the data sets, showing a stable performance across multiple data sets. 
We think IF-PCA has a great potential for single-cell clustering, for the method is simple, transparent, and tuning-free. Although the current IF-PCA(X) still underperforms the state-of-the-art methods (e.g., SC3) in some data sets, it is hopeful that a variant of IF-PCA (say, by borrowing the consensus voting in SC3 or replacing PCA with some other embedding methods \citep{cai2022theoretical, ma2023spectral} can outperform them.

We also find that unsupervised deep learning algorithms do not immediately yield improvements over classical methods on the microarray data and the single-cell data. IF-VAE underperforms IF-PCA in most data sets; there are only a few data sets in which IF-VAE slightly improves IF-PCA. The reason can be either that nonlinear dimension reduction has no significant advantage over linear dimension reduction in these data sets or IF-VAE is not optimally tuned. How to tune the deep learning algorithms in unsupervised settings is an interesting future research direction. 
Moreover, the theory on VAE 
remains largely unknown \citep{FanMa}. A theoretical investigation of VAE requires an understanding to both the deep neural network structures and the variational inference procedure. We also leave this to future work.

The framework of IF-PCA only assumes feature sparsity but no other particular structures on the features. It is possible that the features are grouped \citep{chang2017comparing} or have some tree structures \citep{li2021distance}. How to adapt IF-PCA to this setting is an interesting yet open research direction.

In the real data analysis, we assume that the number of clusters, $K$, is given. 
When $K$ is unknown, how to estimate $K$ is a problem of independent interest.  
One approach is to use the scree plot. For example, \cite{KeMA} proposed a method that first computes a threshold from the bulk eigenvalues in the scree plot and then applies this threshold on the top eigenvalues to estimate $K$. 
Another approach is based on global testing. 
Given a candidate $K$, we may first apply a clustering method with this given $K$ and then apply the global testing methods in \cite{JinKeWang} to test if each estimated cluster has no sub-clusters; $\hat{K}$ is set as the smallest $K$ such that the global null hypothesis is accepted in all estimated clusters.
In general, estimating $K$ is an independent problem from clustering. It is interesting to investigate which estimators of $K$ work best for gene microarray data and single-cell RNA-seq data, which we leave to future work

\end{document}